\documentclass[10pt,twocolumn]{IEEEtran} 

\usepackage{easyReview}
\usepackage{theorem} \usepackage{cite} \usepackage{stfloats} \usepackage{epsfig} \usepackage{verbatim}
\usepackage{graphicx} \usepackage{amssymb} \usepackage{amsmath} \usepackage{color}
\usepackage{bm}
\usepackage{xcolor}
\usepackage{algorithm}
\usepackage{algorithmicx}

\usepackage{multirow} 
\usepackage{siunitx}
\usepackage{booktabs}
\usepackage{graphicx} 
\usepackage{booktabs} 
\usepackage{array}
\usepackage{makecell}
\usepackage{cellspace}
\usepackage{tabularx}
\newcolumntype{L}[1]{>{\raggedright\arraybackslash}p{#1}}
\newcolumntype{C}[1]{>{\centering\arraybackslash}p{#1}}
\newcolumntype{R}[1]{>{\raggedleft\arraybackslash}p{#1}}

\usepackage{graphicx}
\usepackage{subfigure}
\usepackage{float}
\usepackage{placeins}
\usepackage{subfig}
\usepackage{overpic}
\usepackage{epstopdf}
\usepackage{CJKutf8}
\usepackage{indentfirst}
\usepackage{amsmath}
\usepackage{enumitem}
\usepackage{textcomp}
\usepackage{gensymb}
\usepackage{makecell}
\usepackage{subfigure}
\usepackage{subcaption}

\usepackage{ifthen}

\usepackage{titlesec}

\newboolean{insertCitation}
\setboolean{insertCitation}{true} 

\newcommand{\conditionalCite}[1]{%
    \ifthenelse{\boolean{insertCitation}}{\cite{#1}}{}%
}

\begin{document}
\begin{CJK}{UTF8}{gbsn}

\def\QEDclosed{\mbox{\rule[0pt]{1.3ex}{1.3ex}}}
\def\QEDopen{{\setlength{\fboxsep}{0pt}\setlength{\fboxrule}{0.2pt}\fbox{\rule[0pt]{0pt}{1.3ex}\rule[0pt]{1.3ex}{0pt}}}}
\def\QED{\QEDopen}
\def\proof{}
\def\endproof{\hspace*{\fill}~\QED\par\endtrivlist\unskip}

\title{A Superdirective Beamforming Approach based on MultiTransUNet-GAN}
\author{Yali~Zhang,
    Haifan~Yin,~\IEEEmembership{Senior Member,~IEEE},
    and Liangcheng~Han
\thanks{
Yali~Zhang, Haifan~Yin and Liangcheng~Han are with School of Electronic Information and Communications, Huazhong University of Science and Technology, 430074 Wuhan, China (email: yalizhang@hust.edu.cn; yin@hust.edu.cn; hanlc@hust.edu.cn).
}
\thanks{The corresponding author is Haifan~Yin.
}
\thanks{
This work was supported by the National Natural Science Foundation of China under Grant 62071191. 
}
}

\maketitle

\begin{abstract}
In traditional multiple-input multiple-output (MIMO) communication systems, the antenna spacing is often no smaller than half a wavelength. However, by exploiting the coupling between more closely-spaced antennas, a superdirective array may achieve a much higher beamforming gain than traditional MIMO. In this paper, we present a novel utilization of neural networks in the context of superdirective arrays. Specifically, a new model called MultiTransUNet-GAN is proposed, which aims to forecast the excitation coefficients to achieve ``superdirectivity" or ``super-gain" in the compact uniform linear or planar antenna arrays. In this model, we integrate a multi-level guided attention and a multi-scale skip connection. Furthermore, generative adversarial networks are integrated into our model. To improve the prediction accuracy and convergence speed of our model, we introduce the warm up aided cosine learning rate (LR) schedule during the model training, and the objective function is improved by incorporating the normalized mean squared error (NMSE) between the generated value and the actual value. Simulations demonstrate that the array directivity and array gain achieved by our model exhibit a strong agreement with the theoretical values. Overall, it shows the advantage of enhanced precision over the existing models, and a reduced requirement for measurement and the computation of the excitation coefficients. 

\end{abstract}




\section{Introduction}\label{sec:introduction}
Over the years, extensive research and development have been conducted on Multiple-Input Multiple-Output (MIMO), which greatly improves the spectral and energy efficiencies of wireless communication systems. In the commercial antenna arrays like massive MIMO, the antenna spacing is usually set to no less than half a wavelength. While this reduces the coupling effect between antennas, thereby simplifying the complexity of signal reception and transmission in the antenna array, it also imposes limitations on the number of antenna elements within a fixed aperture, which in turn affects the performance of the antenna array, such as the directivity, gain, spectral efficiency. In recent years, there has been a growing interest in enhancing the performance of antenna arrays, leading to an increase in research on dense antenna arrays \cite{pizzo2020spatially} and superdirective antenna arrays \cite{han2022coupling, dovelos2023superdirective, dovelos2022superdirective, tariq2020design, han2020characteristic}. The work \cite{pizzo2020spatially} has highlighted the potential benefits of utilizing holographic MIMO technology to deploy antenna panels on base stations in a seamless and space-efficient manner. As the number of antennas on a fixed panel increases, the interaction between antennas intensifies, necessitating consideration of the coupling effect.\par

To leverage the coupling effect between antennas, it is essential to comprehend the underlying mechanism responsible for this effect. Researches indicate that the external field distribution of an antenna array can be delineated as a superposition of plane waves, which are composed of propagating waves and evanescent waves. The coupling effect between antennas is also influenced by the presence of evanescent waves \cite{clemmow2013plane}. Theoretically, the directivity of the antenna array can be significantly enhanced by exploiting the coupling effect between antennas. When the antenna spacing decreases towards zero and there is precise control over the amplitude and phase of each antenna excitation, the directivity of the array can increase proportionally to the square of the number of antennas. This type of antenna array configuration is commonly known as the ``superdirective array" \cite{uzkov1946approach,altshuler2005monopole}. In contrast, conventional arrays can only achieve directivity proportional to the number of antennas in the array. Nevertheless, precisely solving the beamforming vectors (i.e. excitation coefficients) to maximize the directivity of superdirective arrays poses a significant challenge in practical applications \cite{altshuler2005monopole}.\par

To date, there exist several approaches for determining the excitation coefficients of superdirective arrays. These methods include array theory-based methods \cite{altshuler2005monopole}, circuit theory-based methods \cite{IvrlacTowardAC}, spherical wave expansion methods \cite{clemente2015design}, and coupling model-based methods \cite{han2022coupling}. These methods necessitate intricate analytical deduction prior to determining the excitation coefficients. Additionally, techniques based on array theory and circuit theory mainly rely on theoretical principles, while in practical antenna arrays, the complicated coupling effect is hard to characterize by theory accurately. Applying the excitation coefficients derived from these two methods directly to the antenna arrays can lead to pattern distortion, resulting in a significant discrepancy between the achieved directivity and the theoretical value. The methods based on spherical wave expansion and coupling matrix account for the coupling effect, yielding directivity results that align well with theoretical values. Nevertheless, both approaches require tedious full-wave simulations or measurements. In particular, the coupling matrix-based methods mandate solving the coupling matrix between antennas before determining the excitation coefficients. Overall, these two methods incur high measurement and computation costs, and their implementation is complex for real-world applications.\par

The coupling matrix-based method \cite{han2022coupling} reveals that the solution of the excitation coefficients primarily hinges on the radiated electric field of the antenna array. By utilizing the inherent characteristics of neural networks as a ``black box", it becomes feasible to establish a direct relationship between the radiated electric field distribution and the excitation coefficients of the antenna array and implicitly model the coupling effect between antennas (i.e., coupling matrix). To address the limitation of the conventional approaches for determining the excitation coefficients, we propose the utilization of deep learning techniques in the determination of the excitation coefficients for superdirective antenna arrays. More precisely, we propose a novel neural network model, herein referred to as MultiTransUNet-GAN, specifically designed for this task. The utilization of neural network-based method has potential to decrease the expenses associated with measurement and computation required for predicting the excitation coefficients. \par

Our proposed model comprises of two components: a generator and a discriminator. The generator, known as MultiTransUNet, provides an improvement of the U-Net architecture. The discriminator is an ordinary convolutional neural network. The U-Net \cite{unet} architecture was proposed as a solution for biomedical image segmentation, offering high accuracy in this task. One of its key features is the use of skip connections \cite{he2016deep}, which effectively reduce the loss of data information. However, U-Net has two known limitations. Firstly, it struggles to effectively model long-range dependencies and establish connections between data due to the inherent limitations of convolution operations. Secondly, the skip connection mechanism in U-Net is only applied to feature maps of the same scale, resulting in a certain degree of information loss. These limitations are effectively addressed in our proposed MultiTransUNet model through the incorporation of a multi-level guided attention module and a multi-scale skip connection. The multi-level guided attention mechanism consists of a Transformer \cite{vaswani2017attention,transformer} module and a global spatial attention (GSA) module \cite{woo2018cbam}. Additionally, the multi-scale skip connection consists of the skip connection (i.e., the residual connection) and the dense connection \cite{huang2017densely}. The dense connection is integrated into the decoding pathway of the U-Net architecture. In order to improve the accuracy of predictions, we have included the normalized mean squared error (NMSE) between the actual value and the predicted value in the objective function of our model. \par

To train our model, we utilize the data sets consisting of the electric field and corresponding excitation coefficients of the array at varying antenna spacings and different directions. Subsequently, we use the trained model to predict the excitation coefficients for a given direction. To validate the effectiveness of our proposed model, we design a four printed dipole array operating at a frequency of 1.6 GHz. Simulation results demonstrate that the directivity obtained by our model matches well with the theoretical value. In comparison to previous methods employed to resolve superdirective beamforming vectors, our proposed model demonstrates enhanced precision and a reduced cost of measurement and computation.\par

Furthermore, we show that our proposed model has the capability to calculate the excitation coefficients to maximize the array gain when the antenna loss is considered. It is also applicable in the case of uniform planar array. Simulation results demonstrate the efficacy of our model in precisely predicting the excitation coefficients for uniform linear or planar antenna arrays.\par

The main contributions of this paper are as follows:
\begin{itemize}
    \item To the best of our knowledge, this is the first paper to employ deep learning techniques to address the issue of determining the excitation coefficients for achieving ``superdirectivity" and ``super-gain" of compact uniform linear arrays and uniform planar arrays. In contrast to conventional approaches, our model demonstrates superior predictive accuracy while demanding reduced costs for electric field measurement and calculation.

    \item We propose a novel network architecture named MultiTransUNet-GAN, which incorporates the generative adversarial mechanism. The generator (i.e., the MultiTransUNet) of our model integrates a multi-level guided attention module and a multi-scale skip connection. Simulation results demonstrate that our model outperforms the TransUNet network in terms of predictive accuracy. Furthermore, in comparison to the model that exclusively employs the proposed MultiTransUNet, our model exhibits superior predictive accuracy, with the increased complexity of our model being almost negligible.
    
    \item We propose a new objective function in our model, incorporating the minmax game between the generator and discriminator to optimize the log-likelihood for estimating the conditional probability $P\left( Y=y|x \right)$ and the NMSE between the generated and real excitation coefficients. Here, $Y$ denotes whether $x$ represents the true excitation coefficient (with $y=1$) or the generated excitation coefficient (with $y=0$). The inclusion of NMSE helps to improve the optimization capacity of the neural network and facilitate the generator to accurately produce the excitation coefficients.
   
    \item We introduce the warm up aided cosine LR scheduler within the training process of the proposed model, which greatly improves the prediction accuracy.

\end{itemize}\par

\emph{Notation:} Matrices and vectors are represented by boldface uppercase letters and boldface lowercase letters, respectively. $\mathbb{R}^n$ and $\mathbb{C}^n$ denote the n-dimension real and complex number, respectively. $\left( \cdot \right) ^*,\left( \cdot \right) ^{\text{T}}$, $\left( \cdot \right) ^{\text{H}}$ and $\left( \cdot \right) ^{\dag}$ represent the conjugate, the transpose, the conjugate transpose and pseudo-inverse, respectively. $\left| \cdot \right|$ is the absolute value, $\lVert \cdot \rVert $ represents the Euclidean norm, and $\mathbb{E}\left[ \cdot \right] $ represents the expectation operator. 

\section{System Model}
\label{system_modeling} 
We consider an antenna array consisting of $M$ elements. In this study, a far-field approximation is considered. In accordance with the electromagnetic theory, the electric fields in the spherical coordinate system $\left(r, \theta, \phi\right)$ can be obtained:
\begin{equation}
    \label{eq:1}
    \vec{E}\left( \theta ,\phi \right) =k\sqrt{\eta}\sum_{s=1}^2{\sum_{n=1}^{\infty}{\sum_{m=-n}^n{Q_{smn}\vec{K}_{smn}\left( \theta ,\phi \right)}}},
\end{equation}
where $k=\frac{2\pi}{\lambda}$ denotes wavenumber, $\lambda$ is wavelength, $\eta$ is the wave impedance in free space. $Q_{smn}$ and $\vec{K}_{smn}\left( \theta ,\phi \right) $ are the spherical wave coefficient and far-field wave function of the antenna array, respectively. $s,m,n$ denote the wave modes.\par

The spherical wave coefficients of the array can be expressed as the aggregate of the spherical wave coefficient of individual antenna within the array, i.e. $Q_{smn}=\sum_{i=1}^M{a_iQ_{smni}}$. $a_i$ and $Q_{smni},i=1,2,\cdots ,M$ represent the excitation coefficient and the spherical wave coefficient of the $i$-th antenna, respectively. Therefore, the directivity of the $M$-element antenna array in the direction of $\left( \theta _0,\phi _0 \right)$ can be expressed as:
\begin{equation}
    \label{eq:2}
    D\left( \theta _0,\phi _0 \right) =\frac{\left| \sum_{i=1}^M{a_i}\sum_{smn}{Q_{smni}\vec{K}_{smn}\left( \theta _0,\phi _0 \right)} \right|^2}{\sum_{smn}{\left| \sum_{i=1}^M{a_i}Q_{smni} \right|^2}},
\end{equation}
where $\sum_{smn}=\sum_{s=1}^{2}{\sum_{n=1}^{\infty}{\sum_{m=-n}^{n}}}$.\par

Based on the result presented in \cite{xie2023genetic}, Eq. (\ref{eq:2}) can be further expressed as:
\begin{equation}
    \label{eq:3}
    D\left( \theta _0,\phi _0 \right) =\frac{\mathbf{a}^{\text{T}}\mathbf{E}_{\theta _0,\phi _0}\mathbf{E}_{\theta _0,\phi _0}^{\text{H}}\mathbf{a}^*}{\mathbf{a}^{\text{T}}\mathbf{E}^{\text{H}}\mathbf{Ea}^*}\cdot c,
\end{equation}
where $c$ is a constant. $\mathbf{a}=\left[ a_1,a_2,\cdots ,a_M \right] ^{\text{T}}$ denotes the beamforming vectors of the array. The full electric field of the antenna array is denoted by
\begin{equation}
    \label{eq:4}
    \mathbf{E}=\left[ \begin{matrix}
	E_1\left( \theta _1,\phi _1 \right)&		E_2\left( \theta _1,\phi _1 \right)&		\cdots&		E_M\left( \theta _1,\phi _1 \right)\\
	E_1\left( \theta _1,\phi _2 \right)&		E_2\left( \theta _1,\phi _2 \right)&		\cdots&		E_M\left( \theta _1,\phi _2 \right)\\
	\vdots&		\vdots&		\ddots&		\vdots\\
	E_1\left( \theta _l,\phi _q \right)&		E_2\left( \theta _l,\phi _q \right)&		\cdots&		E_M\left( \theta _l,\phi _q \right)\\
\end{matrix} \right] \in \mathbb{C}^{lq\times M},
\end{equation}
where $E_i\left( \theta _k,\phi _f \right) ,\theta _k\in \left[ 0^{\circ},180^{\circ} \right] ,\phi _f\in \left[ 0^{\circ},360^{\circ} \right), k=1,2,\dots,l, f=1,2,\dots,q$ represents the electric field radiated of the $i$-th antenna in the direction of $\left( \theta _k, \phi _f\right)$. The full electric field of each antenna is sampled in both elevation angle $\theta$ and the azimuth angle $\phi$. $l$ and $q$ denote the total number of samples taken in the direction of $\theta$ and $\phi$, respectively.

The electric field radiated by the antenna array in the direction $\left( \theta _0, \phi _0\right)$ is given by
\begin{equation}
    \label{eq:5}
    \mathbf{E}_{\theta _0,\phi _0}=\left[ E_1\left( \theta _0,\phi _0 \right) ,E_2\left( \theta _0,\phi _0 \right) ,\cdots ,E_M\left( \theta _0,\phi _0 \right) \right] ^{\text{T}} \in \mathbb{C}^{M\times 1}
\end{equation}
\par
Eq. (\ref{eq:3}) represents a Rayleigh entropy form, thus the beamforming vectors maximizing the directivity can be determined through eigenvector decomposition, i.e.,
\begin{equation}
    \label{eq:6}
    \left( \mathbf{E}^{\text{H}}\mathbf{E} \right) ^{-1}\left( \mathbf{E}_{\theta _0,\phi _0}\mathbf{E}_{\theta _0,\phi _0}^{\text{H}} \right) \mathbf{a}_0=\kappa_0 \mathbf{a}_0,
\end{equation}
where $\kappa _0$ is the maximum eigenvalue of the decomposition, $\mathbf{a}_0$ denotes the corresponding eigenvector. The excitation coefficients for achieving the maximum directivity towards the direction $\left( \theta _0, \phi _0 \right)$ is  $\mathbf{a}_0$.
The corresponding maximum directivity in the direction of $\left(\theta_0,\phi_0\right)$ is
\begin{equation}
    \label{eq:7}
    D_{max}\left( \theta _0,\phi _0 \right) =\frac{\mathbf{a}_{0}^{\text{T}}\mathbf{E}_{\theta _0,\phi _0}\mathbf{E}_{\theta _0,\phi _0}^{\text{H}}\mathbf{a}_{0}^{*}}{\mathbf{a}_{0}^{\text{T}}\mathbf{E}^{\text{H}}\mathbf{Ea}_{0}^{*}}\cdot c.
\end{equation}

As indicated by Eq. (\ref{eq:6}), the determination of the excitation coefficients is solely dependent on the radiated electric field of the antenna array. As a result, we utilize the ``black box " characteristics of the neural network to compute the excitation coefficients. The input of the neural network is the radiated electric field distribution of the antenna array in the direction of $\left(\theta,\phi\right)$, and the output is the excitation coefficients that maximize the directivity in that direction. Following the determination of the excitation coefficients, the corresponding directivity can be verified through calculation or measurement. The detailed explanation of the structure, operational mechanisms, and training process of the neural network will be presented in Sec. \ref{sec:network}.\par

\section{Superdirective beamforming based on MultiTransUNet-GAN model}
\label{sec:network}
In this section, we introduce the proposed MultiTransUNet-GAN model. The essential element of our model is the Generative Adversarial Networks (GAN) \cite{creswell2018generative}. The generator network, referred to as MultiTransUNet, is trained to establish the correlation between the electric field of the antenna array in various directions and the excitation coefficients that maximize directivity in those directions. The discriminator network is a prevalent convolutional neural network utilized to differentiate between the actual excitation coefficients and the excitation coefficients produced by the generator, thereby aiding in the training of the generator. In subsequent analyses, we denote the generator network and the discriminator network using the symbols $\text{GN}$ and $\text{DN}$, respectively.\par
In order to effectively train our neural network model, it is necessary to break down the electric field data and the excitation coefficients, and to redefine the structure and format of the data. To prevent confusion with symbols used in Sec. II , starting from this section, the electric field of the antenna array will be denoted by $\mathbf{\epsilon}$, while the corresponding excitation coefficients will be represented as $\mathbf{B}$. The radiated electric field of the antenna array in a certain direction is:
\begin{equation}
    \label{eq:8}
     \mathbf{\epsilon}=\left[ \mathbf{e}_1,\mathbf{e}_2,\cdots ,\mathbf{e}_M \right] \in \mathbb{R}^{4\times M}, 
\end{equation}
the corresponding excitation coefficients are :
\begin{equation}
    \label{eq:9}
    \mathbf{B}=\left[ \mathbf{b}_1,\mathbf{b}_2,\cdots ,\mathbf{b}_M \right] \in \mathbb{R}^{2\times M},
\end{equation}
where $\mathbf{e}_i=\left[ \theta ,\phi ,e_{ia},e_{ip} \right] ^{\text{T}} \in \mathbb{R}^{4\times 1}$ and $\mathbf{b}_i=\left[ b_{ia},b_{ip} \right] ^{\text{T}} \in \mathbb{R}^{2\times 1}$ represent the radiated electric field and the excitation coefficients of the $i$-th antenna in this direction, respectively. $\theta$ and $\phi$ indicate the elevation and azimuth angles, respectively. $e_{ia}$ and $e_{ip}$ denote the amplitude and phase of electric field for the $i$-th antenna, respectively. Similarly, $b_{ia}$ and $b_{ip}$ represent the amplitude and phase of excitation coefficients on the $i$-th antenna, respectively.

\begin{figure*}[t]
    \centering
\includegraphics[scale=0.50]{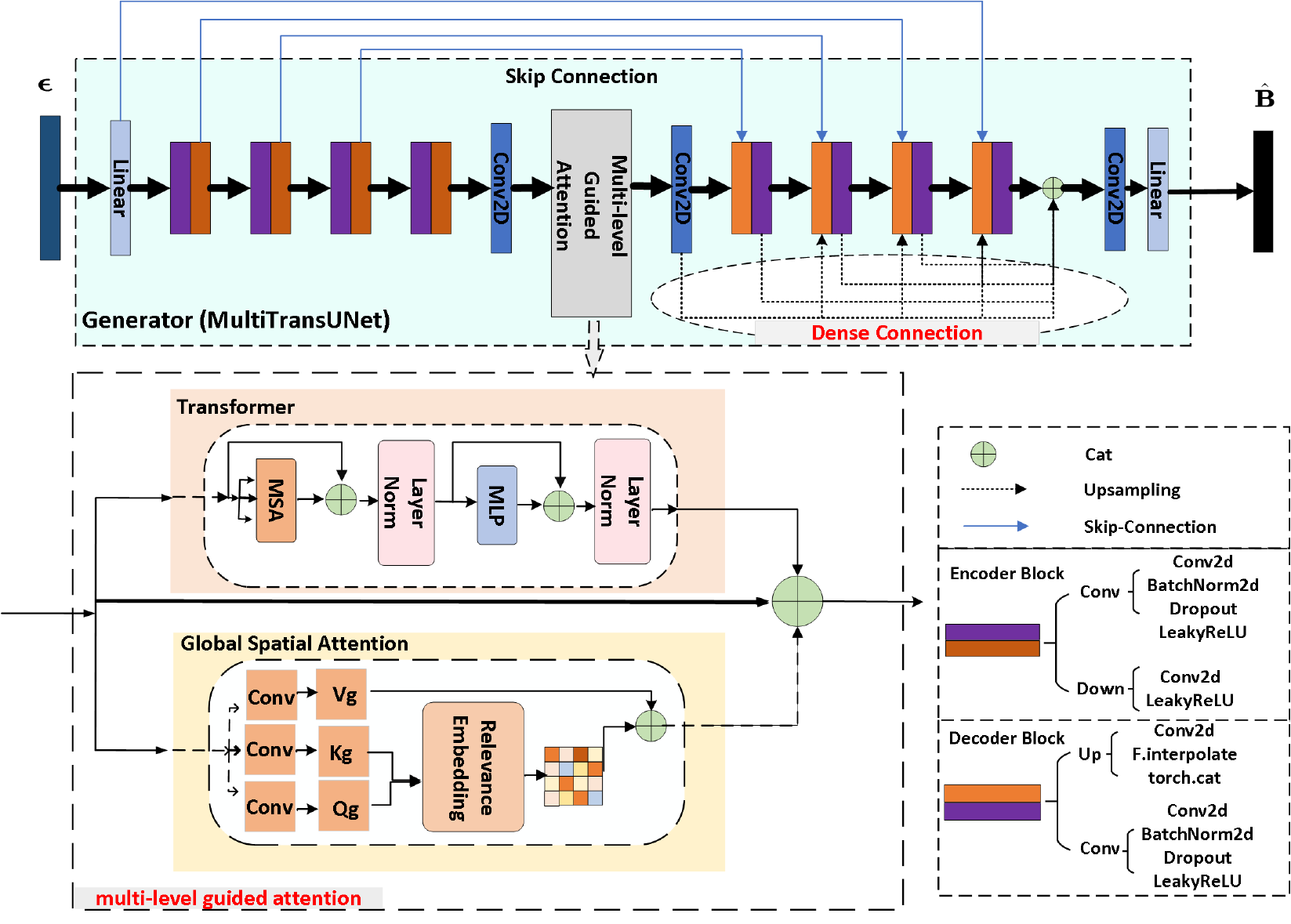}
    \caption{The architecture of the generator based on the proposed MultiTransUNet.}
    \label{fig:generator}
\end{figure*}

\subsection{Objective function}
\label{sec:objectivefunction}
The primary goal of GAN is to improve the capacity of generator to produce realistic data through the game of maximum and minimum confrontation between the generator and the discriminator. However, conventional GAN encounters challenges related to convergence and training instability \cite{towards2017}. In conventional GAN, the data prediction process of generator relies solely on the feedback of discriminator. While this feedback theoretically guides the generator towards improvement through iterative processes, it often fails to provide specific insights such as why the current data is deemed insufficiently authentic or how the generator should adjust its training to enhance the realism of the generated data. Consequently, the feedback information provided by the discriminator in traditional GAN is quite restricted.\par
In an effort to mitigate the problems associated with the traditional GAN, a novel approach has been proposed which incorporates the real excitation coefficients $\mathbf{B}$ as a conditional parameter during the training process of GAN \cite{MirzaConditionalGA,dong2020channel}. This inclusion enables the generator to produce the excitation coefficients in a manner that aligns with the actual data. As a result, the discriminator will only deem the generated data as accurate when it closely resembles the real data, which significantly improves the accuracy of the prediction of the generator. The objective function of our model is thus
\begin{equation}
    \label{eq:10}
    \underset{\varTheta_g}{\min}\underset{\varTheta_d}{\ \max}\ \mathcal{L}_{\text{GAN}}\left( \text{GN}_{\varTheta_g},\text{DN}_{\varTheta_d},\mathbf{\epsilon ,B} \right) +\mathcal{L}_2.
\end{equation}
where $\varTheta_g$ and $\varTheta_d$ represent the learnable parameter of the generator and discriminator network, respectively.\par
$\mathcal{L}_{\text{GAN}}$ denotes the objective function of the traditional GAN 
\begin{equation}
    \label{eq:11}
    \begin{aligned}
        \mathcal{L}_{\text{GAN}}\left( \text{GN}_{\varTheta_g},\text{DN}_{\varTheta  _d},\mathbf{\epsilon ,B} \right) &=\mathbb{E}\left[ \log\text{DN}_{\varTheta _d}\left( \mathbf{B} \right) \right] \\
        &+\mathbb{E}\left[ \log \left( 1-\text{DN}_{\varTheta  _d}\left( \text{GN}_{\varTheta  _g}\left( \mathbf{\epsilon } \right) \right) \right) \right] ,
    \end{aligned}
\end{equation}
which ensures that the generated data closely aligns with the distribution of the actual data. Nevertheless, it lacks the ability to control the specific characteristics of the generated data, thus failing to ensure a close resemblance to the actual data. To break this constraint, we introduce an additional auxiliary condition, specifically the NMSE between the generated data and the authentic data $\mathcal{L}_2$:
\begin{equation}
    \label{eq:12}
    \mathcal{L}_2=\mathbb{E}\left[ \lVert \mathbf{B}-\text{GN}_{\varTheta  _g}\left( \mathbf{\epsilon } \right) \rVert ^2 \right],
\end{equation}
which guarantees that the data produced by the generator is not only derived from the distribution of the actual data but also closely resembles the real data.\par
As the proposed model is trained on batch data each time, a statistical average is employed to determine the objective function which is expressed as the expectation $\mathbb{E}\left[ \cdot \right] $. This optimization objective is crucial for the GAN to effectively learn intricate data distributions and generate samples of superior quality.\par

\subsection{Network architecture}
\label{network_architecture}
In this subsection, we will introduce the specific network architecture of our model, the approach used for data pre-processing, and the training methodology employed for the model.\par

In the generator network, a novel neural network called MultiTransUNet has been proposed specifically for the task at hand. It improves upon the existing TransUNet \cite{Chen2021TransUNetTM} network by integrating a multi-level guided attention module, which comprises a Transformer \cite{vaswani2017attention,transformer} and a GSA module \cite{woo2018cbam}. This module is inserted between the shrinking path and the extension path of the U-Net \cite{unet} architecture. Additionally, the expansion path of the U-Net incorporates the dense connection \cite{huang2017densely}. Numerical results indicate that the proposed MultiTransUNet outperforms TransUNet in the given task when employing the same scale U-Net.\par 

\begin{figure}[t]
    \centering
    \includegraphics[scale=0.35]{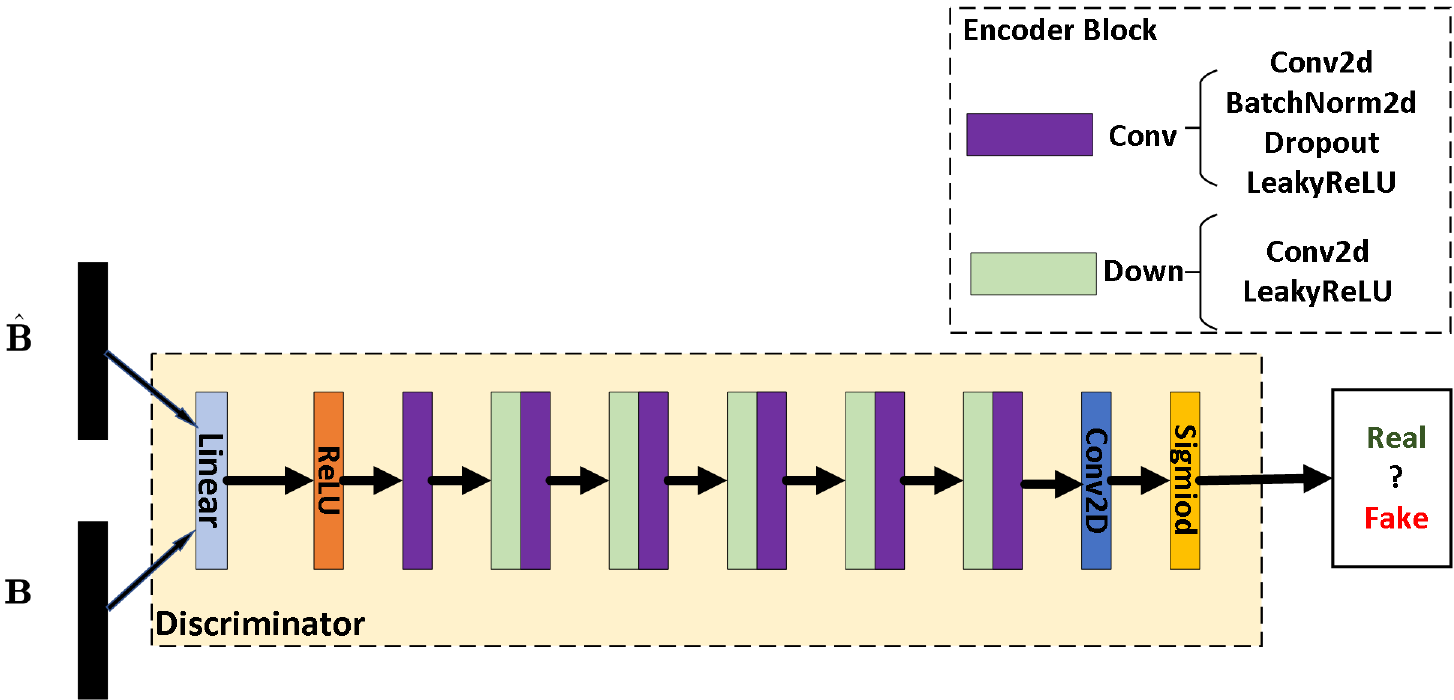}
    \caption{The architecture of the discriminator.}
    \label{fig:discriminator}
\end{figure}

The architectural designs of the generator and discriminator are depicted in Fig. \ref{fig:generator} and Fig. \ref{fig:discriminator}, respectively. Prior to being input into the network, the electric field $\mathbf{\epsilon}$ and the excitation coefficients $\mathbf{B}$ are transformed into matrices with dimensions $M\times 4\times 1$ and $M\times 2\times 1$, respectively. The $i$-th channel of the matrix corresponds to the electric field and the excitation coefficients of the $i$-th antenna, respectively. The second dimension of $\mathbf{\epsilon}$ represents the four components of the electric field of an antenna. Similarly, the second dimension of $\mathbf{B}$ denotes the amplitude and phase of the excitation coefficient of an antenna.\par

For the purpose of feature extraction, the generator employs a series of operations. Initially, the generator scales the size of input data from $M\times 4\times 1$ to $M\times 64\times 64$ using a full connection layer known as Linner. Following this, four encoders are utilized to extract important features from the data. The resulting features are then passed through a multi-level guided attention module, which is subsequently followed by four symmetrical decoders. Prior to entering or exiting the multi-level guided attention module, the data  undergoes a 2-D convolution operation (Conv2d) for reshaping. The decoder is responsible for generating the target data in a step-by-step manner. Finally, the data is processed through a Conv2d layer and a Linner layer to produce the predicted excitation coefficients $\mathbf{\hat{B}}$. Additionally, the skip connections are formed between the encoder and decoder at the same level, while the dense connections are present in the extension path of the U-Net. The feature map of each decoder is upsampled and fed into the subsequent layer. Moreover, the discriminator primarily consists of five encoders. Subsequently, we will present the two components within the multi-level guided attention module.\par

\begin{table}[b]\normalsize
    \centering
    \begin{tabular}{p{8cm}} 
    \toprule 
    \textbf{Algorithm 1:} The Transformer Module\\
    \midrule 
    \textbf{Input:} $\mathbf{X}_t=\left[ \mathbf{x}_{t1},\mathbf{x}_{t2},\cdots ,\mathbf{x}_{tn} \right] $\\
    \textbf{Output:} $\mathbf{O}_t=\left[ \mathbf{o}_{t1},\mathbf{o}_{t2},\cdots ,\mathbf{o}_{tn} \right] $
    \begin{itemize}
        \item [1:] Key matrix in Transformer: $\mathbf{K}_{t}=\mathbf{W}_{t}^k\mathbf{X}_{t}$ 
        \item [2:] Query matrix in Transformer: $\mathbf{Q}_{t}=\mathbf{W}_{t}^q\mathbf{X}_{t}$
        \item [3:] Value matrix in Transformer: $\mathbf{V}_{t}=\mathbf{W}_{t}^{v}\mathbf{X}_{t}$
        \item [4:] Attention matrix in Transformer: 
                   \begin{equation}
                       \mathbf{Y}_t=\text{Softmax} \left( \frac{\mathbf{K}_{t}^{\text{T}}\mathbf{Q}_t}{\sqrt{d}} \right)\nonumber
                   \end{equation}
        \item [5:] $\mathbf{Z}_t=\text{LN}\left( \mathbf{X}_t+\mathbf{V}_t\mathbf{Y}_t \right)$
        \item [6:] $\mathbf{O}_t=\text{LN}\left( \mathbf{Z}_t+\text{FCN}\left( \mathbf{Z}_t \right) \right) $
    \end{itemize} \\
    \bottomrule 
    \end{tabular}
\end{table}

\subsubsection{Transformer }The procedure of the Transformer module is shown in \textbf{Algorithm 1}. The encoded feature is represented by $\mathbf{X}_{en}\in \mathbb{R}^{c\times w\times h}$, where $c$, $w$ and $h$ are the number of channel, width and height of $\mathbf{X}_{en}$, respectively. For the subsequent utilization of the Transformer module, the encoded feature is transformed into $\mathbf{X}_t=\left[ \mathbf{x}_{t1},\mathbf{x}_{t2},\cdots ,\mathbf{x}_{tn} \right] \in \mathbb{R}^{m\times n}$, where $n$ denotes the length of the encoded data channels and $m$ denotes the size of feature map for each channel, i.e., $n=c$ and $m=w\times h$ . We perform different linear transformations on $\mathbf{X}_{t}$:

\begin{equation}
    \label{eq:13}
    \mathbf{K}_{t}=\mathbf{W}_{t}^k\mathbf{X}_{t},
\end{equation}
\begin{equation}
    \label{eq:14}
    \mathbf{Q}_{t}=\mathbf{W}_{t}^q\mathbf{X}_{t},
\end{equation}
\begin{equation}
    \label{eq:15}
    \mathbf{V}_{t}=\mathbf{W}_{t}^{v}\mathbf{X}_{t},
\end{equation}
where $\mathbf{K}_t=\left[ \mathbf{k}_{t1},\mathbf{k}_{t2},\cdots ,\mathbf{k}_{tn} \right] \in \mathbb{R}^{d\times n}$, $\mathbf{Q}_t=\left[ \mathbf{q}_{t1},\mathbf{q}_{t2},\cdots ,\mathbf{q}_{tn} \right] \in \mathbb{R}^{d\times n}$ and $\mathbf{V}_t=\left[ \mathbf{v}_{t1},\mathbf{v}_{t2},\cdots ,\mathbf{v}_{tn} \right] \in \mathbb{R}^{m\times n}$ are the key matrix, the query matrix and the value matrix, respectively; $\mathbf{W}_{t}^{k}\in \mathbb{R}^{d\times m}$, $\mathbf{W}_{t}^{q}\in \mathbb{R}^{d\times m}$ and $\mathbf{W}_{t}^{v}\in \mathbb{R}^{m\times m}$ are the corresponding trainable linear transformation matrix, respectively; $d$ is the feature dimension of the column vector in the matrix $\mathbf{K}_{t}$ and generally set to be consistent with the input dimension $m$. Specially, the key matrix $\mathbf{K}_{t}$ and the query matrix $\mathbf{Q}_{t}$ are utilized for the purpose of executing the weight assignment function. $\mathbf{K}_{t}^{\text{T}}\mathbf{Q}$ represents the correlation between input and output data of this module.\par

Based on $\mathbf{K}_{t}$ and $\mathbf{Q}_{t}$, we can obtain the attention matrix $\mathbf{Y}_{t}$:
\begin{equation}
        \label{eq:16}
        \mathbf{Y}_t=\text{Softmax} \left( \frac{\mathbf{K}_{t}^{\text{T}}\mathbf{Q}_t}{\sqrt{d}} \right),
\end{equation}
where the softmax operation $\text{Softmax} \left( \mathbf{x} \right) =\frac{\exp \left( x_i \right)}{\sum{\exp \left( x_i \right)}}$ is used for normalization.

To avoid the issue of diminishing gradients resulting from data instability, we use a residual connection \cite{he2016deep} and layer normalization in the subsequent step. 
\begin{equation}
        \label{eq:17}
        \mathbf{Z}_t=\text{LN}\left( \mathbf{X}_t+\mathbf{V}_t\mathbf{Y}_t \right),
\end{equation}
where $\text{LN}\left( \mathbf{X} \right) =\gamma \frac{x_{ij}-\mu _j}{\sqrt{\sigma _{j}^{2}+\zeta}}+\beta $ denotes layer normalization, in which $\mu_{j}$, $\sigma_{j}$ are the mean and standard deviation of the $j$-th column of matrix $\mathbf{X}$. Scale $\gamma$ and bias vector $\beta$ are parameters of the layer normalization. $\zeta$ is added to improve the numerical stability of denominator and can be set to $1\times 10^{-5}$. Finally, a two-layer FCN is used to extract features further, which could be written as:
\begin{equation}
        \label{eq:18}
        \mathbf{O}_t=\text{LN}\left( \mathbf{Z}_t+\text{FCN}\left( \mathbf{Z}_t \right) \right),
\end{equation}
where $\mathbf{O}_t=\left[ \mathbf{o}_{t1},\mathbf{o}_{t2},\cdots ,\mathbf{o}_{tn} \right] \in \mathbb{R}^{m\times n}$ denotes the output of the Transformer module, and $\text{FCN}\left( \mathbf{X} \right) =\mathbf{W}_{t}^{2}\max \left( 0,\mathbf{W}_{t}^{1}\mathbf{X}+\mathbf{u}_{t}^{1} \right) +\mathbf{u}_{t}^{2}$, $ \mathbf{W}_{t}^{1}\in \mathbb{R}^{d_m\times m}$, $\mathbf{u}_{t}^{1}\in \mathbb{R}^{d_m\times 1}$, $\mathbf{W}_{t}^{2}\in \mathbb{R}^{m\times d_m}$, $\mathbf{u}_{t}^{2}\in \mathbb{R}^{m\times 1}$, in which $d_m$ is the length of the hidden layer in the FCN and $d_m$ is set to $4m$. Moreover, we reshape the resulting features to obtain the final output of the Transformer module, which is denoted by $\mathbf{F}_t\in \mathbb{R}^{c\times w\times h}$.\par

\begin{table}[t]\normalsize
    \centering
    \begin{tabular}{p{8cm}} 
    \toprule 
    \textbf{Algorithm 2:} The Global Spatial Attention Module \\
    \midrule 
    \textbf{Input:} $\mathbf{X}_g$\\
    \textbf{Output:} $\mathbf{O}_g$
    \begin{itemize}
        \item [1:] Key matrix in GSA: $\mathbf{K}_g=M_k\left( \mathbf{X}_g \right) $ 
        \item [2:] Query matrix in GSA: $\mathbf{Q}_{g}=M_q\left( \mathbf{X}_g \right)$
        \item [3:] Value matrix in GSA: $\mathbf{V}_{g}=M_v\left( \mathbf{X}_g \right) $
        \item [4:] Attention matrix in GSA: 
                   \begin{equation}
                       \mathbf{Z}_g=\text{Softmax} \left( \frac{\mathbf{K}_{g}^{\text{T}}\mathbf{Q}_g}{\sqrt{d}} \right)\nonumber
                   \end{equation}
        \item [5:] $\mathbf{O}_g=\mathbf{V}_g\mathbf{Z}_g$
    \end{itemize} \\
    \bottomrule 
    \end{tabular}
\end{table}

\subsubsection{Global Spatial Attention (GSA) }The procedure of the GSA is shown in \textbf{Algorithm 2}. The input and output of GSA are $\mathbf{X}_g\in \mathbb{R}^{c\times w\times h}$ and $\mathbf{O}_g\in \mathbb{R}^{c\times w\times h}$, respectively. $\mathbf{X}_g=\mathbf{X}_{en}$. Firstly, the input $\mathbf{X}_{g}$ undergoes three separate operations: $M_{k}$, $M_{q}$, and $M_{v}$. Each operation involves a 2D convolution and reshaping procedures, resulting in generating three feature maps: $\mathbf{K}_g\in \mathbb{R}^{c^{\prime }\times \left( w\times h \right)}$, $\mathbf{Q}_g\in \mathbb{R}^{c^{\prime }\times \left( w\times h \right)}$ and $\mathbf{V}_g\in \mathbb{R}^{c\times \left( w\times h \right)}$, where $c^{\prime}=c/\gamma$ and $\gamma=8$. Next, matrix multiplication is performed with $\mathbf{K}_{g}$ and $\mathbf{Q}_{g}$, which is followed by the softmax normalization.

\begin{equation}
    \label{eq:19}
    \mathbf{Z}_g=\text{Softmax} \left( \frac{\mathbf{K}_{g}^{\text{T}}\mathbf{Q}_g}{\sqrt{d^{'}}} \right),
\end{equation}
where $d^{\prime}=w\times h$ denotes the length of the column vector of the matrix $\mathbf{K}_{g}$. After normalization, $\mathbf{V}_{g}$ is multiplied with $\mathbf{Z}_g$, and the resulting feature can be formulated as
\begin{equation}
    \label{eq:20}
    \mathbf{O}_g=\mathbf{V}_g\mathbf{Z}_g,
\end{equation}
where $\mathbf{O}_g\in \mathbb{R}^{c\times \left( h\times w \right)}$. Moreover, we reshape the resulting features to obtain the final output of GSA, which is denoted by $\mathbf{F}_g\in \mathbb{R}^{c\times w\times h}$.

In the multi-level guided attention module, the feature map is combined using weighted methods to optimize the utilization of available data information. The resulting output feature of the multi-level guided attention is represented as $\mathbf{F}_a$, as below:
\begin{equation}
    \label{eq:21}
    \mathbf{F}_a=\mathbf{X}_{en}+\mathbf{F}_t+\mathbf{F}_g,
\end{equation}
where $\mathbf{X}_{en}$, $\mathbf{F}_t$ and $\mathbf{F}_g$ represent the encoded feature, the feature map of the output of the Transformer module and the Global Spatial Attention module, respectively.\par

\begin{table}[t]\normalsize
    \centering
    \caption{MultiTransUNet Hyperparameters}
    \resizebox{\linewidth}{!}{\begin{tabular}{c|c|c|c}
    \hline
    \hline
    \multicolumn{3}{c|}{\textbf{Layer Name}}& \textbf{Configuration}  \\
    \hline
    \hline
    \multicolumn{3}{c|}{Input}     &  $M\times 4\times 1$  \\
    \hline
    \multicolumn{3}{c|}{Linear}    &  $\left(M\times 4\times 1, M\times 64\times 64\right)$ \\
    \hline
    \multirow{2}{*}{\text{Encoder\ Block\ }1} & \multicolumn{2}{c|}{Conv} & $\left( 4,\ 64,\ 3,\ 1,\ 1 \right) $\\
    \cline{2-4}
     & \multicolumn{2}{c|}{Down} &  $\left( 64,\ 64,\ 3,\ 2,\ 1 \right) $ \\
     \hline
     \multirow{2}{*}{\text{Encoder\ Block\ }2} & \multicolumn{2}{c|}{Conv} & $\left( 64,\ 128,\ 3,\ 1,\ 1 \right) $\\
    \cline{2-4}
     & \multicolumn{2}{c|}{Down} &  $\left( 128,\ 128,\ 3,\ 2,\ 1 \right) $ \\
     \hline
     \multirow{2}{*}{\text{Encoder\ Block\ }3} & \multicolumn{2}{c|}{Conv} & $\left( 128,\ 256,\ 3,\ 1,\ 1 \right) $\\
    \cline{2-4}
     & \multicolumn{2}{c|}{Down} &  $\left( 256,\ 256,\ 3,\ 2,\ 1 \right) $ \\
     \hline
     \multirow{2}{*}{\text{Encoder\ Block\ }4} & \multicolumn{2}{c|}{Conv} & $\left( 256,\ 512,\ 3,\ 1,\ 1 \right) $\\
    \cline{2-4}
     & \multicolumn{2}{c|}{Down} &  $\left( 512,\ 512,\ 3,\ 2,\ 1 \right) $ \\
     \hline
     \multicolumn{3}{c|}{Conv2d} & $\left( 512,\ 768,\ 1,\ 1,\ 0 \right) $ \\
     \hline
     \multirow{2}{*}{Transformer} & \multicolumn{2}{c|}{MSA}& $\text{heads}=8$ \\
     \cline{2-4}
     & \multicolumn{2}{c|}{MLP} & $\left( 768,\ 3072,\ 768 \right) $ \\
     \hline
     \multicolumn{3}{c|}{GSA} & $\gamma =8$ \\
     \hline
     \multicolumn{3}{c|}{Conv2d} & $\left( 768,\ 512 ,\ 1,\ 1,\ 0 \right) $ \\
     \hline
     \multirow{3}{*}{\text{Decoder\ Block\ }1} & \multirow{2}{*}{Up} & F.interpolate & $\text{scale\_factor}=2$\\
     \cline{3-4}
      & & Conv2d & $\left( 512 ,\ 256 ,\ 1,\ 1,\ 0 \right) $\\
     \cline{2-4}
      & \multicolumn{2}{c|}{Conv2d} & $\left( 256\times 2,\ 256,\ 1,\ 1,\ 0 \right) $ \\
     \hline
     \multirow{3}{*}{\text{Decoder\ Block\ }2} & \multirow{2}{*}{Up} & F.interpolate & $\text{scale\_factor}=2$\\
     \cline{3-4}
      & & Conv2d & $\left( 256 ,\ 128 ,\ 1,\ 1,\ 0 \right) $\\
     \cline{2-4}
      & \multicolumn{2}{c|}{Conv2d} & $\left( 128\times 3,\ 128,\ 1,\ 1,\ 0 \right) $ \\
     \hline
     \multirow{3}{*}{\text{Decoder\ Block\ }3} & \multirow{2}{*}{Up} & F.interpolate & $\text{scale\_factor}=2$\\
     \cline{3-4}
      & & Conv2d & $\left( 128 ,\ 64 ,\ 1,\ 1,\ 0 \right) $\\
     \cline{2-4}
      & \multicolumn{2}{c|}{Conv2d} & $\left( 64\times 4,\ 64,\ 1,\ 1,\ 0 \right) $ \\
     \hline
     \multirow{3}{*}{\text{Decoder\ Block\ }4} & \multirow{2}{*}{Up} & F.interpolate & $\text{scale\_factor}=2$\\
     \cline{3-4}
      & & Conv2d & $\left( 64 ,\ 32 ,\ 1,\ 1,\ 0 \right) $\\
     \cline{2-4}
      & \multicolumn{2}{c|}{Conv2d} & $\left( 32\times 4+3, \ 32,\ 1,\ 1,\ 0 \right) $ \\
     \hline
     \multicolumn{3}{c|}{Conv2d}  & $\left( 32\times 5,\ M,\ 3,\ 1,\ 1 \right) $  \\
     \hline
     \multicolumn{3}{c|}{Linear} & $\left( M\times 64\times 64,\ M\times 4\times 1 \right) $\\
     \hline
     \multicolumn{3}{c|}{Output} & $M\times 2\times 1$   \\
     \hline
    \hline
    \end{tabular}}
    \label{tab:I}
\end{table}
The configuration of the primary layers within the MultiTransUNet model are presented in Table I. The network configuration of Conv block is denoted by $\left( C_{in},\ C_{out},\ K,\ S,\ P \right) $, in which $C_{in}$ and $C_{out}$ correspond to the number of channels of the input and output feature, respectively. The parameter $K$ signifies the size of the convolution kernel, while $S$ indicates the sampling interval of the convolution kernel as it traverses the input feature map. $P$ represents the fill value used in the convolution process. The network configuration of Linear is denoted by $\left(I, O\right)$, in which $I$ and $O$ represent the dimension of input and output feature maps, respectively. The configuration of MLP in the Transformer module is denoted by $\left( L_{in},\ L_{mid},\ L_{out} \right) $, in which $L_{in}$, $L_{mid}$ and $L_{out}$ represent the length of the input, hidden, and output layers, respectively.\par 

\begin{table}[t]\normalsize
    \centering
    \begin{tabular}{p{8cm}} 
    \toprule 
    \textbf{Algorithm 3:} {Superdirective Beamforming Vectors Prediction Method based on the Proposed MultiTransUNet-GAN Model} \\
    \midrule 
    \textbf{Input:} {Electric field distribution $\mathbf{\epsilon}$, the excitation coefficients $\mathbf{B}$}\\
    \textbf{Output:} {Predicted excitation coefficients $\mathbf{\hat{B}}$}\\
    \textbf{Initialization:} {Learning rate of generator $\alpha_g$, learning rate of discriminator $\alpha_d$, exponential decay rate of moment estimates of generator $\beta _g$, exponential decay rate of moment estimates of discriminator $\beta _d$, batch size of the train sets $m_1$, batch size of the test sets $m_2$, antenna array size $M\times 1$}\\
    \textbf{Process:}\\
    \textbf{For} $\text{epoch}=1,2,\cdots ,\mathcal{K}$ \textbf{do}\\
    \quad \textbf{For} $\text{step}=1,2,\cdots ,K$ \textbf{do}
    \begin{itemize}
        \item Sample $m_1$ electric field samples $\left\{ \mathbf{\epsilon}^1,\mathbf{\epsilon}^2,\cdots ,\mathbf{\epsilon}^{m_1} \right\} $
        \item Sample $m_1$ actual excitation coefficients samples $\left\{ \mathbf{B}^1,\mathbf{B}^2,\cdots ,\mathbf{B}^{m_1} \right\} $
        \item Update discriminator by using adaptive moment estimation:
         \begin{equation}
            \begin{split}
                \nabla _{\theta _d}\frac{1}{m_1}\sum_{i=1}^{m_1}
                \left\{ \log \left( 1-\text{DN}\left( \text{GN}\left( \mathbf{\epsilon}^i \right) \right) \right)\right.\\
                \left. +\log \text{DN}\left( \mathbf{B}^i \right)  \right\} \nonumber
            \end{split}
        \end{equation}
        \item Update generator by using adaptive moment estimation:
        \begin{equation}
            \begin{split}
              \nabla _{\theta _g}\frac{1}{m_1}\sum_{i=1}^{m_1} 
                 \left\{\log \left( 1-\text{DN}\left( \text{GN}\left( \mathbf{\epsilon}^i \right) \right) \right)\right.\\ 
                 \left. +\lVert \mathbf{B}^i-\text{GN}\left( \mathbf{\epsilon}^i \right) \rVert ^2 \right\} \nonumber
            \end{split}
        \end{equation}
    \end{itemize}\\
    \quad \textbf{End for}\\
    \textbf{End for}\\
    \bottomrule 
    \end{tabular}
\end{table}

Prior to training the model, data pre-processing is conducted. In order to improve the speed at which the model converges and prevent numerical bi-polarization in the input data, we implemented Min-Max normalization on the amplitude and phase of the excitation coefficients. This normalization method is demonstrated as follows:
\begin{equation}
    \label{eq:22}
    \left\{ \begin{array}{l}
	\hat{b}_{ia}=\frac{b_{ia}-\min \left( b_{ia} \right)}{\max \left( b_{ia} \right) -\min \left( b_{ia} \right)}\\
	\hat{b}_{ip}=\frac{b_{ip}-\min \left( b_{ip} \right)}{\max \left( b_{ip} \right) -\min \left( b_{ip} \right)}\\
\end{array} \right..
\end{equation}

\begin{figure}[h]
    \centering
    \includegraphics[scale=0.30]{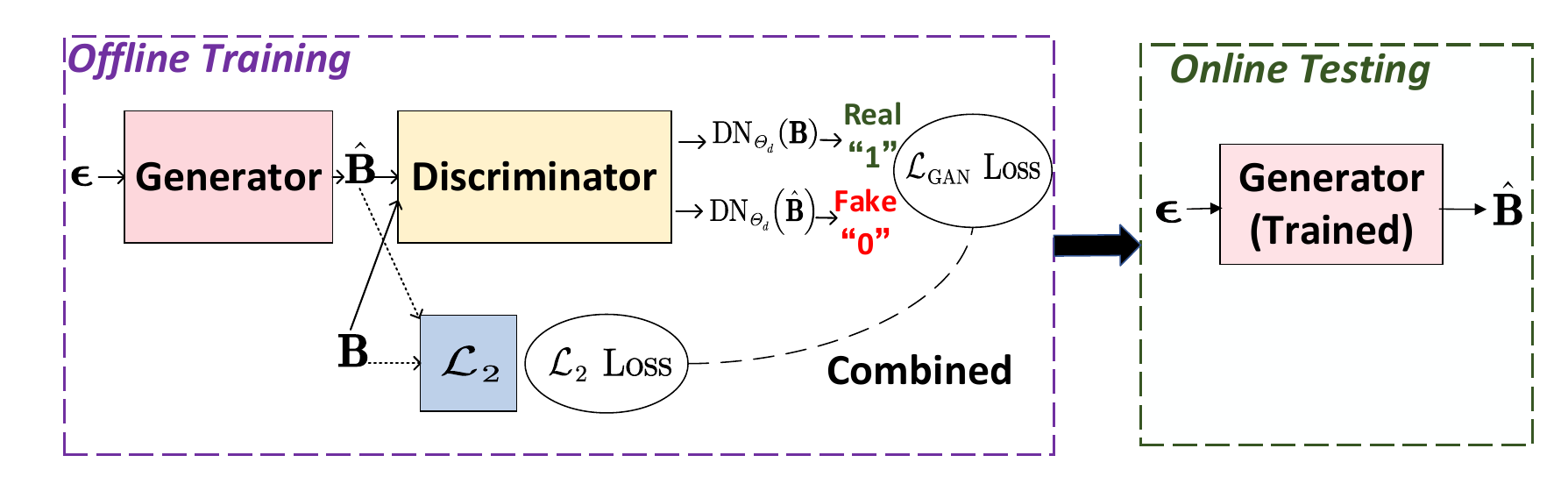}
    \caption{Training block diagram of the proposed MultiTransUNet-GAN model.}
    \label{fig:training-diagram}
\end{figure}
During the model training process, we first conduct offline training followed by online testing, as illustrated in Fig. \ref{fig:training-diagram}. The pseudocode for the algorithm predicting excitation coefficients based on MultiTransUNet-GAN is outlined in \textbf{Algorithm 3}.

\subsection{Complexity Analysis}
\label{complexity_analysis}
In this subsection, we evaluate the complexity of the proposed MultiTransUNet-GAN model in terms of the number of multiplications. Given that the complexity of discriminator is significantly lower than that of the generator, we use the complexity of the generator as a representation of the overall complexity of our model.\par

The generator model primarily consists of the encoder module, the decoder module, and the multi-level guided attention module. We first analyze the complexity of the multi-level guided attention module. According to \textbf{Algorithm 1}, we observe that the complexity of the Transformer module mainly comes from the attention mechanism from step 1 to 5 and the FCN processing in step 6. In the first three steps, we need to perform the computation of the product between the input of the Transformer module $\mathbf{X}_t\in \mathbb{R}^{m\times n}$ and the respective linear transformation matrices $\mathbf{W}_{t}^{k}\in \mathbb{R}^{d\times m}$, $\mathbf{W}_{t}^{q}\in \mathbb{R}^{d\times m}$, $\mathbf{W}_{t}^{v}\in \mathbb{R}^{m\times m}$. These operations have the complexity in the order of $\mathcal{O}\left( mnd \right)$, $\mathcal{O}\left( mnd \right)$ and $\mathcal{O}\left( m^2n \right)$, respectively. Similarity, step 4 and step 5 have the complexity in the order of $\mathcal{O}\left( n^2d \right)$ and $\mathcal{O}\left( n^2m \right) $, respectively. In step 6, the complexity of the multiplication between $\mathbf{Z}_{t}$ and $\mathbf{W}_{t}^{i}$ is $\mathcal{O}\left( mnd_m \right) $. Thus, the computational complexity of the Transformer module is in the order of $\mathcal{O}\left( m^2n \right) +\mathcal{O}\left( n^2m \right)$. According to \textbf{Algorithm 2}, step 1 to step 3 have the complexity in the order of $\mathcal{O}\left( whcc^{\prime} \right) $, $\mathcal{O}\left( whcc^{\prime} \right) $ and $\mathcal{O}\left( whc^2 \right)$, respectively. Similarity, step 4 and step 5 have the complexity in the order of $\mathcal{O}\left( w^2h^2c^{\prime}\right)$ and $\mathcal{O}\left( w^2h^2c \right)$,  respectively. Due to $c^{\prime}=c/8$, the computational complexity of the GSA module is in the order of $\mathcal{O}\left( whc^2 \right) +\mathcal{O}\left( w^2h^2c \right) $. In general, the multi-level guided attention module has a complexity order of $\mathcal{O}\left( whc^2 \right) +\mathcal{O}\left( w^2h^2c \right)$.

Next, we proceed to determine the computational complexity associated with the encoding and decoding component of the generator model. The encoding part consists of four encoder blocks and a Linear layer. Symmetrically, the decoding part consists of four decoder blocks, a Linear layer and a Conv2d layer. In the process of encoding, it is assumed that the feature map has a maximum width of $W_m$, a maximum length of $H_m$ and a maximum number of channels of $C_m$. According to Table I, the encoding part and the decoding part of the generator model have the complexity in the order of $\mathcal{O}\left( W_mH_mC_{m}^{2} \right) $ and $\mathcal{O}\left( W_mH_mC_{m}^{2} \right) $ , respectively.\par
In general, the overall complexity of the generator model is in the order of $\mathcal{O}\left( W_mH_mC_{m}^{2} \right)  +\mathcal{O}\left( whc^2 \right) +\mathcal{O}\left( w^2h^2c \right) $, in which $c=2C_m,\ w=W_m/16,\ h=H_m/16$. Therefore, the overall complexity of the generator model can be summarized as $\mathcal{O}\left( W_mH_mC_{m}^{2} \right) $.

\section{Numerical results and analysis}
\label{sec:numericalResult}

In this section, we describe the methodology used to obtain the training and testing sets for our model. We evaluate the prediction capabilities of our proposed model, MultiTransUNet-GAN, on the superdirective beamforming vectors and compare its performance with that of our proposed MultiTransUNet model and traditional TransUNet model. It is worth noting that the traditional TransUNet model mentioned in this research refers to the TransUNet model that incorporates a single Transformer module \cite{Chen2021TransUNetTM}.\par

\begin{figure}[h]
    \centering
    \includegraphics[scale=0.65]{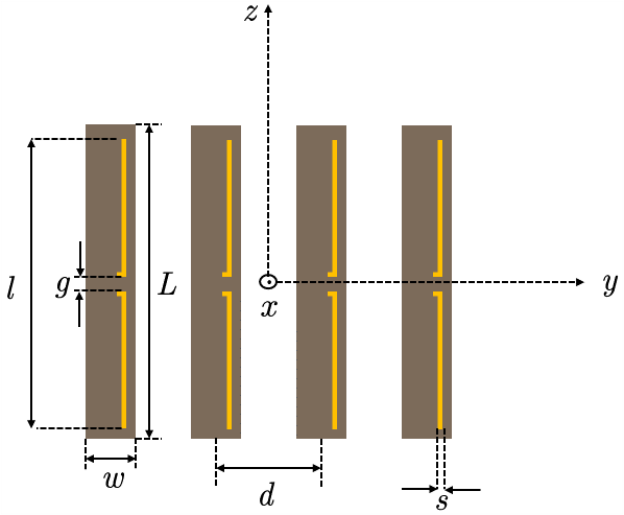}
    \caption{The antenna array of four printed dipole antennas.}
    \label{fig:4-element_array}
\end{figure}

First, we introduce the detailed design of an antenna array comprising four printed dipole antennas, as depicted in Fig. \ref{fig:4-element_array}, which are modeled through the utilization of a high-frequency simulation software, CST. The antenna parameters in the model are as follows: the antenna is an ideal dipole with the length of $l=74.04\ mm$ and the arm spacing of $g=2.54\ mm$. The antenna width is $s=1.0\ mm$. The media substrate uses FR-4 material ($\epsilon _r=4.3,\mu _r=1.0,\tan \delta =0.025$). The dielectric substrate has a width of $w=12.2\ mm$, a height of $L=78\ mm$ and a thickness of $t=0.8\ mm$. Each antenna in the array has a resonant frequency of 1.6 GHz.\par

In this study, the CST software is employed to randomly model antenna arrays with various antenna spacing $d\in \left[ 0.10\lambda ,0.50\lambda \right] $. Initially, we simulate the full radiated electric field when only one antenna in the array is excited by a unit excitation. Subsequently, the electric field is sampled at $5^{\degree}$ intervals in the direction $(\theta,\phi)$, where $\theta \in [0^{\degree}, 180^{\degree}]$, $\phi \in [0^{\degree}, 360^{\degree})$. In general, we can obtain the 2664 sets of electric field in different directions after each sampling. Finally, Eq. (\ref{eq:6}) is employed to calculate the corresponding excitation coefficients for each direction.\par

Based on the aforementioned data acquisition, we obtained 401 sets of the full electric field of the uniform linear array with various antenna spacings. After sampling, the size of datasets $\left\{ \mathbf{\epsilon,B} \right\}$ is 1,068,264. We remove any anomalous data and split the datasets, using $70\%$ for training and $30\%$ for testing. The entire pipeline is implemented using PyTorch. The Adam algorithm is employed to update the parameters. The specific simulation parameters of our model can be found in Table \ref{tab:II}.\par

\begin{table}[t]\normalsize
    \centering
    \caption{Simulation Parameters}
    \begin{tabular}{C{5cm}|C{1.2cm}|C{1.2cm}}
    \hline
         \multicolumn{2}{c|}{\textbf{MultiTransUNet-GAN Parameters}} & \textbf{Value}  \\ 
         \hline
         \multirow{2}{*}{Learning rate}   & $\alpha_g$ & $4\times 10^{-4}$  \\
         \cline{2-3}
         & $\alpha_d$ & $4\times 10^{-5}$ \\
         \hline
         \multirow{2}{*}{Exponential decay rate} & $\beta_g$ & $\left(0.5, 0.9\right)$ \\
         \cline{2-3}
         & $\beta_d$ & $\left(0.5, 0.9\right)$ \\
         \hline
         \multirow{2}{*}{Batch size} & $m_1$ & 200 \\
         \cline{2-3}
         & $m_2$ & 200 \\
         \hline
         Antenna array size & $M\times 1$ & $4\times 1$\\
         \hline
         Number of epochs & $\mathcal{K}$& 50\\
         \hline
         Number of steps in each epoch & $k$& 3739 \\
         \hline
    \end{tabular}
    \label{tab:II}
\end{table}

To assess the performance, the NMSE is employed to quantify the disparity between the original $\mathbf{B}$ and the predicted $\mathbf{\hat{B}}$, as outlined in Eq. (\ref{eq:23}). The prediction precision of the model on the test set is gauged by Acc, as defined in Eq. (\ref{eq:24}).
\begin{equation}
    \label{eq:23}
    \text{NMSE}=10\log _{10}\left\{ \mathbb{E}\left[ \frac{\lVert \mathbf{B}-\mathbf{\hat{B}} \rVert ^2}{\lVert \mathbf{B} \rVert ^2} \right] \right\} \left( \text{dB} \right) ,
\end{equation}
\begin{equation}
    \label{eq:24}
    \text{Acc}=\left( 1-\frac{|\mathbf{B}-\mathbf{\hat{B}|}}{|\mathbf{B|}} \right) \times 100\% .
\end{equation}\par

\begin{figure}
    \centering
    \includegraphics[scale=0.60]{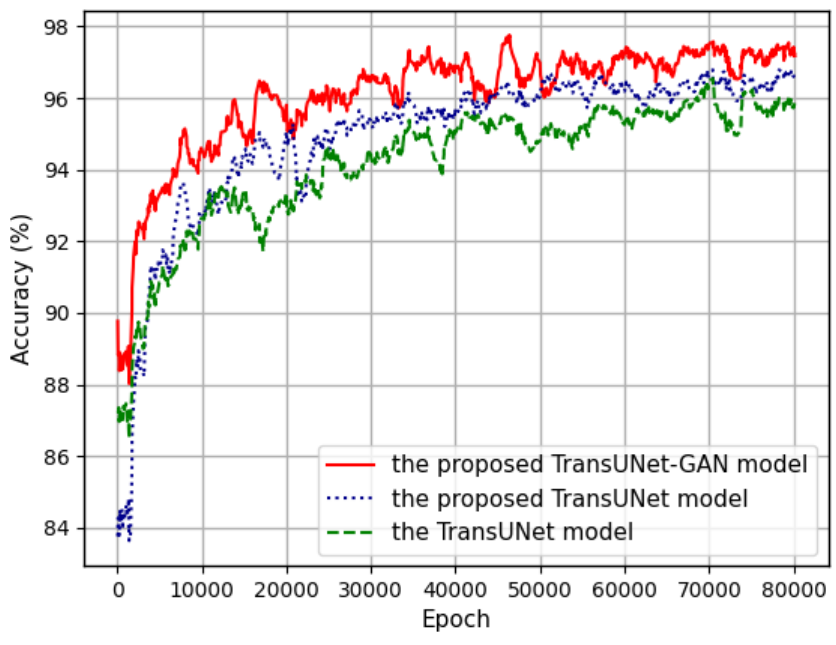}
    \caption{The prediction accuracy of different models}
    \label{fig:acc_transunet}
\end{figure}

\begin{figure}
    \centering
    \includegraphics[scale=0.85]{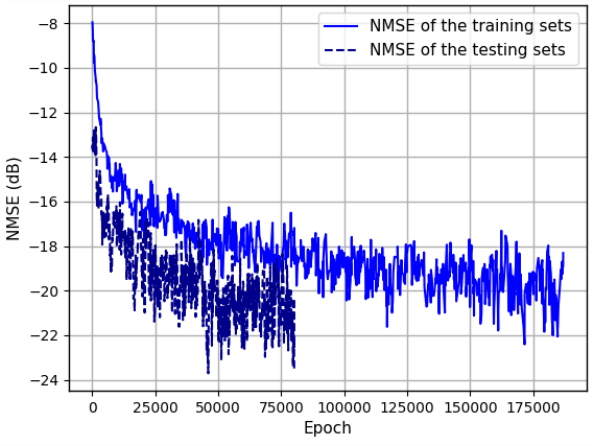}
    \caption{The NMSE of the proposed MultiTransUNet-GAN model vs. the number of epochs.}
    \label{fig:nmse_transunet}
\end{figure}

Next, the prediction performances of three models are compared: the original TransUNet model, the proposed MultiTransUNet model, and our proposed MultiTransUNet-GAN model. Fig. \ref{fig:acc_transunet} demonstrates that all three models are capable of convergence, and the MultiTransUNet-GAN model achieves higher prediction accuracy after the same number of training epochs. The inclusion of the adversarial mechanism aids in the learning process of the neural network, while the utilization of multi-level attention and multi-scale skip connections effectively minimizes information loss during feature extraction. Fig. \ref{fig:nmse_transunet} shows that the NMSE of both the training and testing sets can reach approximately -22dB.\par

\begin{table*}[t]
    \centering
    \caption{The achieved directivity of different methods}
    \begin{tabularx}{\textwidth}{c|c|S|c|S|c}
       \hline
       \textbf{d} & \textbf{array theory-based} & \textbf{Circuit theory-based} &  \textbf{Coupling matrix-based} & \textbf{MultiTransUNet-GAN model-based} & \textbf{Theoretical value}\\
       \hline
        $d=0.15\lambda $ & 6.1090 & 6.6910 & 17.4900 & \textbf{17.4500} & 18.1600\\
        \hline
        $d=0.20\lambda $ & 9.3120 & 9.5560 &16.4800 & \textbf{16.3400} & 17.1400\\ 
        \hline
        $d=0.25\lambda$ & 11.5000 & 11.3600 & 15.2400 & \textbf{15.1900} & 15.7800\\
        \hline
        $d=0.30\lambda$ & 11.8700 & 11.8100 & 13.6700 & \textbf{13.6000} & 14.0800 \\
        \hline
        $d=0.35\lambda$ & 10.8400 & 10.7600 & 11.7500 & \textbf{11.6900} & 12.0100\\
        \hline
        $d=0.40\lambda$ & 8.9490 & 8.8390 & 9.3870 & \textbf{9.3770} & 9.6000 \\
        \hline
    \end{tabularx}
    \label{tab:III}
\end{table*}

To verify the effectiveness of the MultiTransUNet-GAN model, we applied the trained model to predict the excitation coefficients for antenna arrays with varying antenna spacings in diverse directions. The obtained excitation coefficients are then used for simulation.

Table \ref{tab:III} presents the theoretical end-fire directivity and the end-fire directivity obtained from the array theory-based method \cite{altshuler2005monopole}, circuit theory-based method \cite{IvrlacTowardAC}, coupling matrix-based method \cite{han2022coupling}, and MultiTransUNet-GAN method for various antenna spacings in the 4-dipole antenna array. The results in Table \ref{tab:III} indicate that the directivity obtained by the array theory-based method and the circuit theory-based method is less effective for smaller antenna spacing, and this ineffectiveness worsens as the antenna spacing decreases. The diminished performance can be attributed to the absence of accounting for the coupling effect between antennas in both the array theory-based and circuit theory-based methods. In contrast, both the MultiTransUNet-GAN method and the coupling method yield directivity that closely align with the theoretical values.

Nevertheless, by employing the proposed MultiTransTransUNet-GAN model-based approach, the prediction of the excitation coefficients in a specific direction can be achieved by solely considering the radiated electric field in that direction, whose dimension is $M\times 4\times 1$. $M$ denotes the quantity of antennas in the array. The final two dimensions, i.e., `` $4\times 1$ ", signify the radiated electric field of an antenna in that specific direction, encompassing elevation angle $\theta$, azimuth angle $\phi$, and the magnitude and phase of the point electric field. Note that $\theta$ and $\phi$ are known parameters in the specified direction. Conversely, the method relying on the coupling matrix needs to measure the full electric field from each antenna within the array both with and without coupling effect. The dimension of the full electric field is $M\times 4\times 2664$. And the coupling-matrix technique is limited to antenna spacing configurations supported by existing measurement data. In contrast, our model possesses the capacity to forecast the excitation coefficients for optimizing the directivity of the array across various configurations, encompassing the antenna spacings that have not been previously trained. Overall, the method based on MultiTransUNet-GAN offers a simpler approach.\par

\begin{figure}[h]
        \center
        \scriptsize
        \begin{tabular}{cc}
                \includegraphics[scale=0.35]{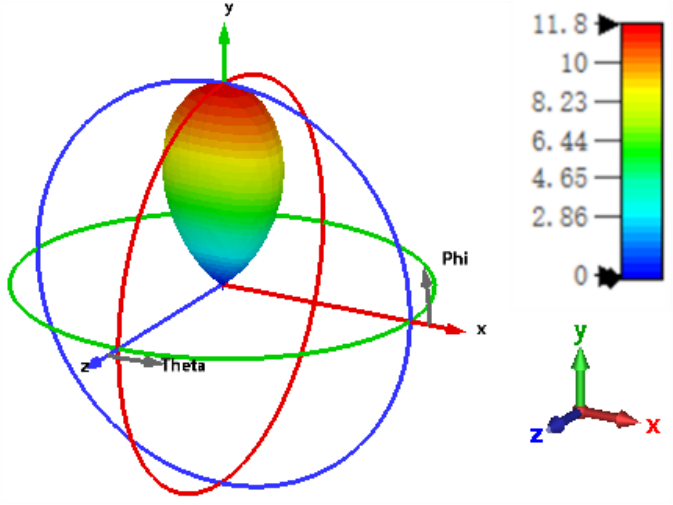}    &     \includegraphics[scale=0.35]{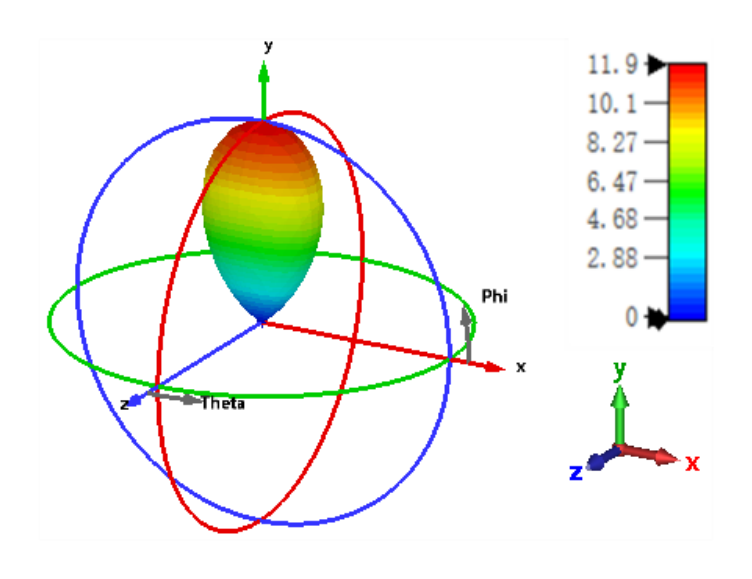} \\
                (a) & (b)
        \end{tabular}
        \caption{When the antenna spacing is $0.30\lambda$, the 3D radiation patterns of the linear dipole antenna array in the direction of $\left( 90^\circ ,60^\circ  \right)$. (a) Based on the MultiTransUNet-GAN model. (b) Based on the coupling matrix-based method. }
        \label{fig:90-60-3d}
        \vspace{-0.5em}
\end{figure}

\begin{figure}[h]
        \center
        \scriptsize
        \begin{tabular}{cc}
                \includegraphics[scale=0.35]{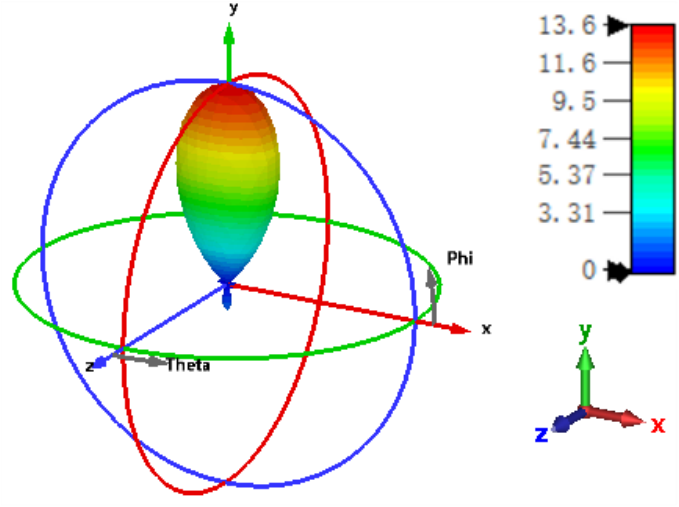}    &     \includegraphics[scale=0.35]{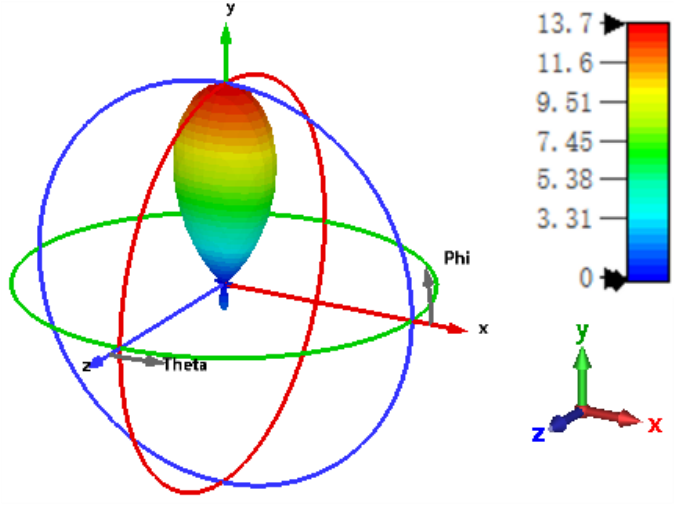} \\
                (a) & (b)
        \end{tabular}
        \caption{When the antenna spacing is $0.30\lambda$, the 3D radiation patterns of the linear dipole antenna array in the direction of $\left( 90^\circ ,80^\circ  \right)$. (a) Based on the MultiTransUNet-GAN model. (b) Based on the coupling matrix-based method.}
        \label{fig:90-80-3d}
        \vspace{-0.5em}
\end{figure}

As depicted in Fig. \ref{fig:4-element_array}, the direction of the end-fire is $\left( 90^{\degree},90^{\degree} \right)$. In addition to our primary investigation, we conducted simulations to evaluate the directivity of the linear dipole antenna array in different orientations. Fig. \ref{fig:90-60-3d} and Fig. \ref{fig:90-80-3d} present the 3D radiation patterns in the directions of $\left( 90^{\degree},60^{\degree} \right)$ and $\left(90^{\degree}, 80^{\degree}\right)$ for the antenna spacing of $0.30\lambda$. The figures illustrate that the radiation patterns in directions aside from end-fire, as derived from both the proposed MultiTransUNet-GAN model and the coupling matrix-based method, also exhibit a satisfactory level of agreement.\par

\begin{table}[b]
    \centering
    \caption{The performance comparison between different models on our datasets}
    \begin{tabular}{|c|c|c|c|}
    \hline
    Method & FLOPs & Params & Accuracy \\
    \hline
     TransUNet & 1488.82 M & 17.82 M  & 95.5\% \\
     \hline
     MultiTransUNet & 2044.76 M & 18.83 M & 96.0\% \\
     \hline
     \textbf{MultiTransUNet-GAN} &  \textbf{2046.54 M} & \textbf{18.99 M} & \textbf{97.6}\%\\
     \hline
     \end{tabular}
    \label{tab:IV}
\end{table}

Moreover, Table \ref{tab:IV} summarizes the FLOPs and the number of parameters of network models, including TransUNet, the proposed MultiTransUNet and MultiTransUNet-GAN model. The total number of FLOPs of the proposed MultiTransUNet-GAN model is about $2046.54\ \text{M}$ and is almost the same as the FLOPs of MultiTransUNet. The MultiTransUNet-GAN model demonstrates a superior average prediction accuracy of 97.6\%, surpassing the prediction accuracy of the MultiTransUNet model by approximately 1.6\%.\par

\begin{figure*}[t]
\centering
\subfigure[]
{
    \begin{minipage}[b]{.31\linewidth}
        \centering
        \includegraphics[scale=0.40]{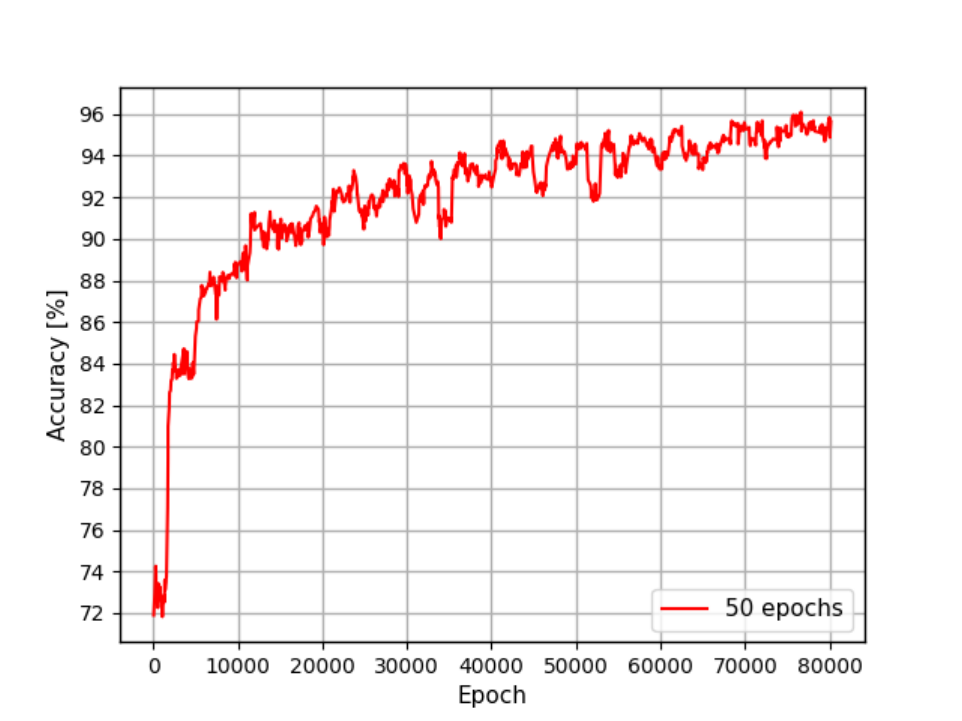}
    \end{minipage}
}
\subfigure[]
{
 	\begin{minipage}[b]{.31\linewidth}
        \centering
        \includegraphics[scale=0.40]{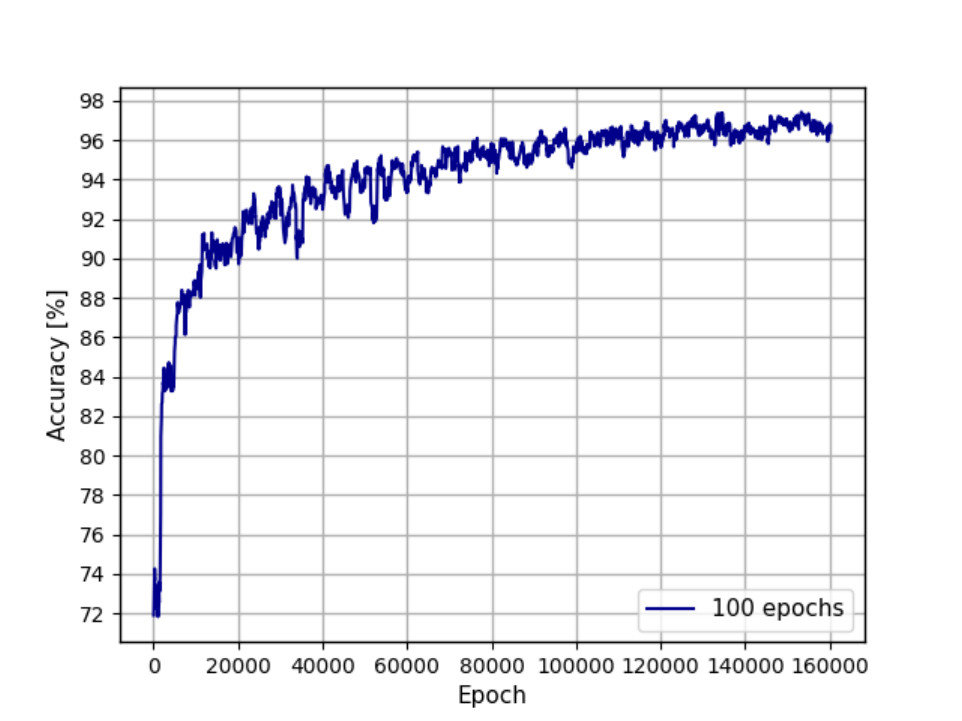}
    \end{minipage}
}
\subfigure[]
{
 	\begin{minipage}[b]{.31\linewidth}
        \centering
        \includegraphics[scale=0.40]{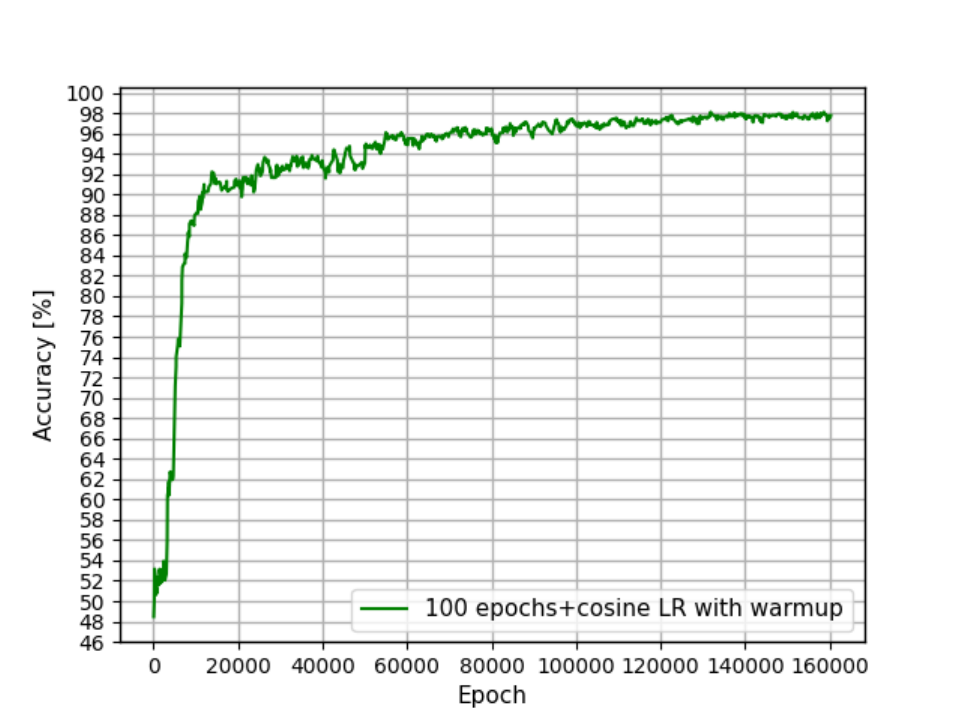}
    \end{minipage}
}
\caption{The accuracy of the ``super-gain" prediction model. (a) Based on 50 training epochs. (b) Based on 100 training epochs. (c) Based on the 100 training epochs and the the warm up aided cosine LR scheduler.}
\label{fig:acc_gain}
\end{figure*}

\begin{figure*}[h]
\centering
\subfigure[]
{
    \begin{minipage}[b]{.31\linewidth}
        \centering
        \includegraphics[scale=0.30]{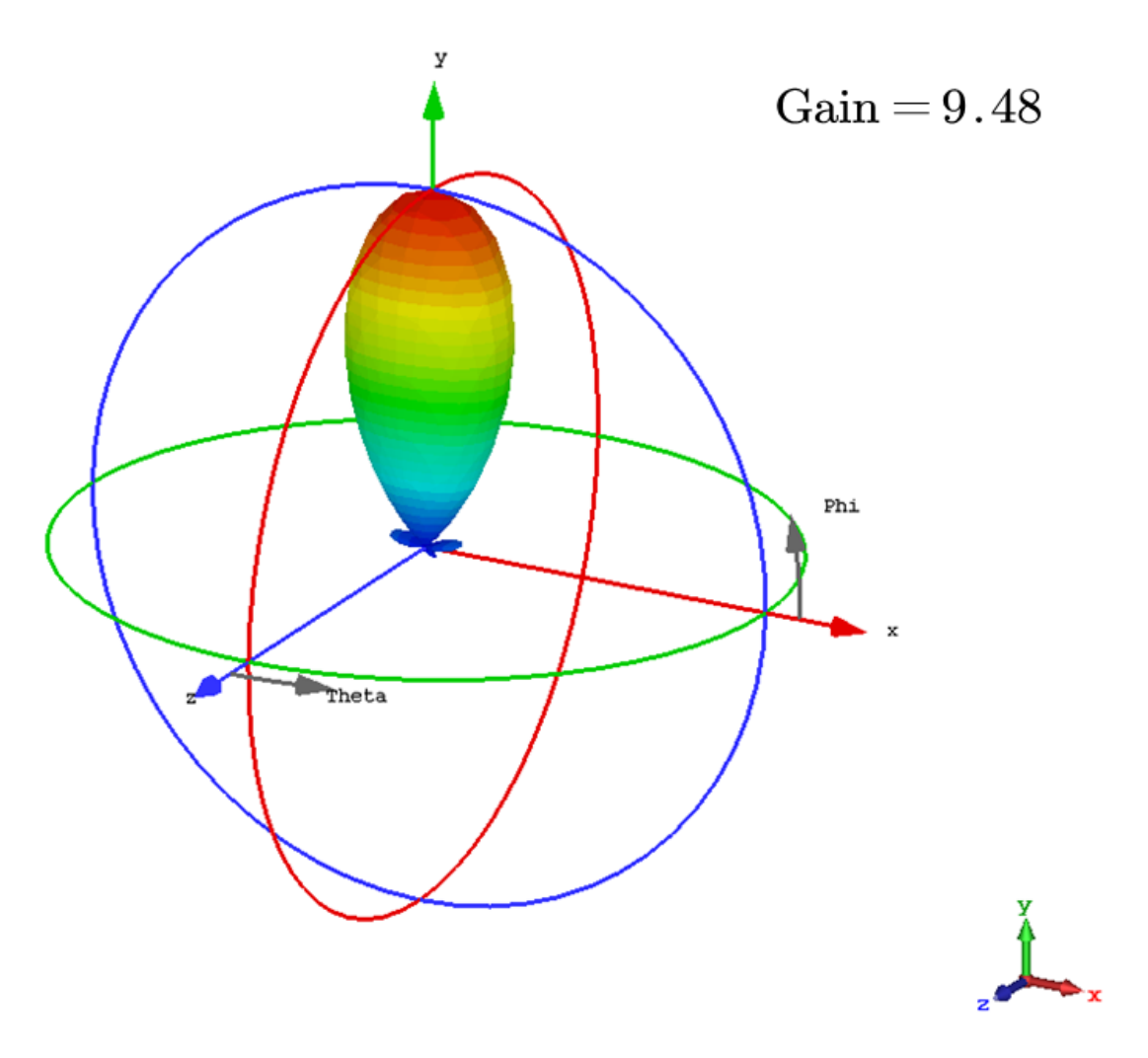}
    \end{minipage}
}
\subfigure[]
{
 	\begin{minipage}[b]{.31\linewidth}
        \centering
        \includegraphics[scale=0.30]{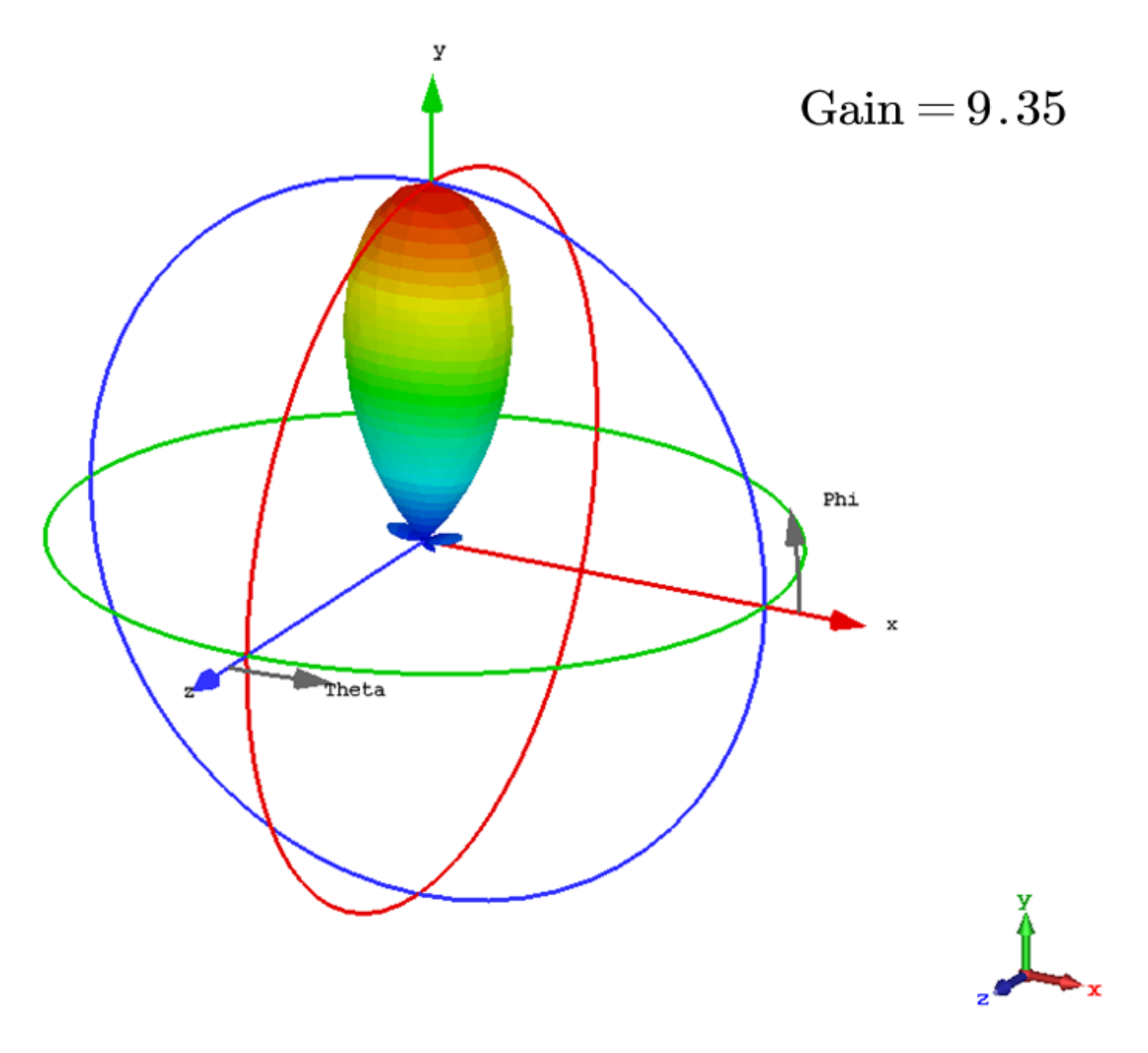}
    \end{minipage}
}
\subfigure[]
{
 	\begin{minipage}[b]{.31\linewidth}
        \centering
        \includegraphics[scale=0.30]{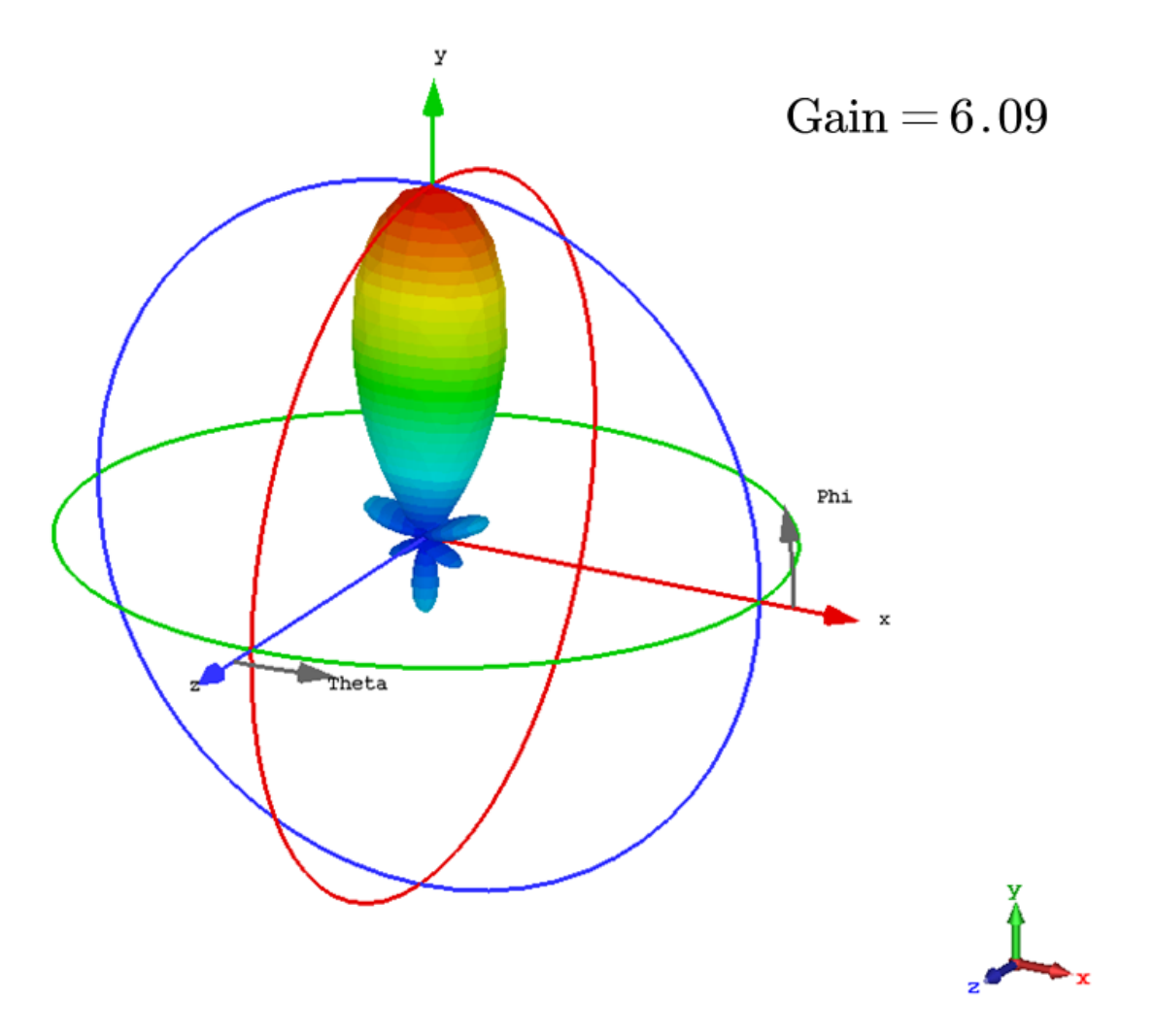}
    \end{minipage}
}
\caption{When the antenna spacing is $0.25\lambda$, the array gain obtained by (a) the proposed MultiTransUNet-GAN model; (b) the coupling matrix-based method; (c) the traditional methods (neglecting the coupling effect).}
\label{fig:pattern_gain}
\end{figure*}

\section{Lossy Antennas -- Array Gain}
\label{sec:array_gain}
In practical application, realistic dipole antennas exhibit a conduction/loss resistance which leads to heat dissipation. Conventional approaches for determining the excitation coefficients to maximize the gain of the dipole antenna array often assume that the dipole is a linear wire of infinitesimal size. Nevertheless, in practical applications, it is not feasible to construct dipole antennas as infinitesimally thin wires. In this study, we utilize CST simulation to obtain the antenna radiation efficiency $\eta$ of the printed dipole. Subsequently, we calculate the loss impedance $r_{loss}$ of the antenna based on Eq. (\ref{eq:25}).
\begin{equation}
    \label{eq:25}
    r_{loss}=\frac{1-\eta}{\eta}.
\end{equation}

\subsection{Data Acquisition and Model Training}
\label{sec:data_accquisition_gain}
We directly utilized the datasets of electric field that had been previously compiled in order to forecast the excitation coefficients for achieving superdirectivity in the 4-element printed dipole array. The CST simulation indicates that the radiation efficiency of the antenna in Fig. \ref{fig:4-element_array} is 0.9546. Subsequently, we determine the excitation coefficients that maximizes the array gain by following the procedure described in \cite{han2022coupling}. Overall, we have obtained a set of training data comprising $401\times 2664$ pairs. The neural network model and the training methodology employed are analogous to those in Sec. \ref{sec:numericalResult}.\par

As illustrated in Fig. \ref{fig:acc_gain}, in order to improve the predictive precision of the model, the training period was extended from 50 to 100 epochs, leading to an accuracy level of approximately $96.5\%$, denoting a $1.5\%$ improvement. Additionally, to further enhance the accuracy of the model, the warm up aided cosine LR scheduler was implemented based on the 100 training epochs and the prediction accuracy of our model reached around $98\%$ as shown in Fig. \ref{fig:acc_gain} (c).\par

In the training process of the ``super-gain" prediction model with the warm up aided cosine LR schedule, the learning rate of the generator $\alpha_{g}$ undergoes a linear increase from $\alpha_{g}^{min}=4\times 10^{-6}$ to $\alpha_{g}^{max}=1\times 10^{-3}$ in the initial 20 epochs, known as ``warm-up", followed by a cosine decrease in the final 80 epochs, as described in Eq. (\ref{eq:28}).
\begin{equation}
    \alpha _g=\alpha _{g}^{min}+\frac{1}{2}\left( \alpha _{g}^{max}-\alpha _{g}^{min} \right) \left( 1+\cos \left( \frac{t-T_{\max}}{T-T_{\max}}\pi \right) \right) ,
    \label{eq:28}
\end{equation}
where $T_{max}$ and $T$ are the number of warm up and total epochs, respectively. This warm up aided cosine annealing algorithm facilitates rapid convergence of the model in the early stages and prevents it from being stuck in local optima due to high learning rates in later stages.\par

\subsection{Numerical results and analysis}
\label{sec:numerical_gain}
Fig. \ref{fig:pattern_gain} shows the 3D end-fire array gain patterns of the array with antenna spacing of $0.25\lambda$ obtained by our proposed model, the coupling matrix-based method \cite{han2022coupling} and the conventional approach (neglecting the coupling effect) \cite{altshuler2005monopole, IvrlacTowardAC}. The comparison reveals that the array gain derived from the conventional approach (neglecting the coupling effect) is 6.09, whereas our proposed model yields a higher gain of 9.48.\par

\section{Planar Array}
\label{sec:planar_array}
In comparison to the linear arrays, planar arrays are more versatile due to the 3D beam steering capabilities. In this section, we have utilized the proposed neural network model to predict the excitation coefficients to maximize the directivity of the $4\times 4$ uniform planar array (UPA).\par

\begin{figure}[h]
    \centering
    \includegraphics[scale=0.30]{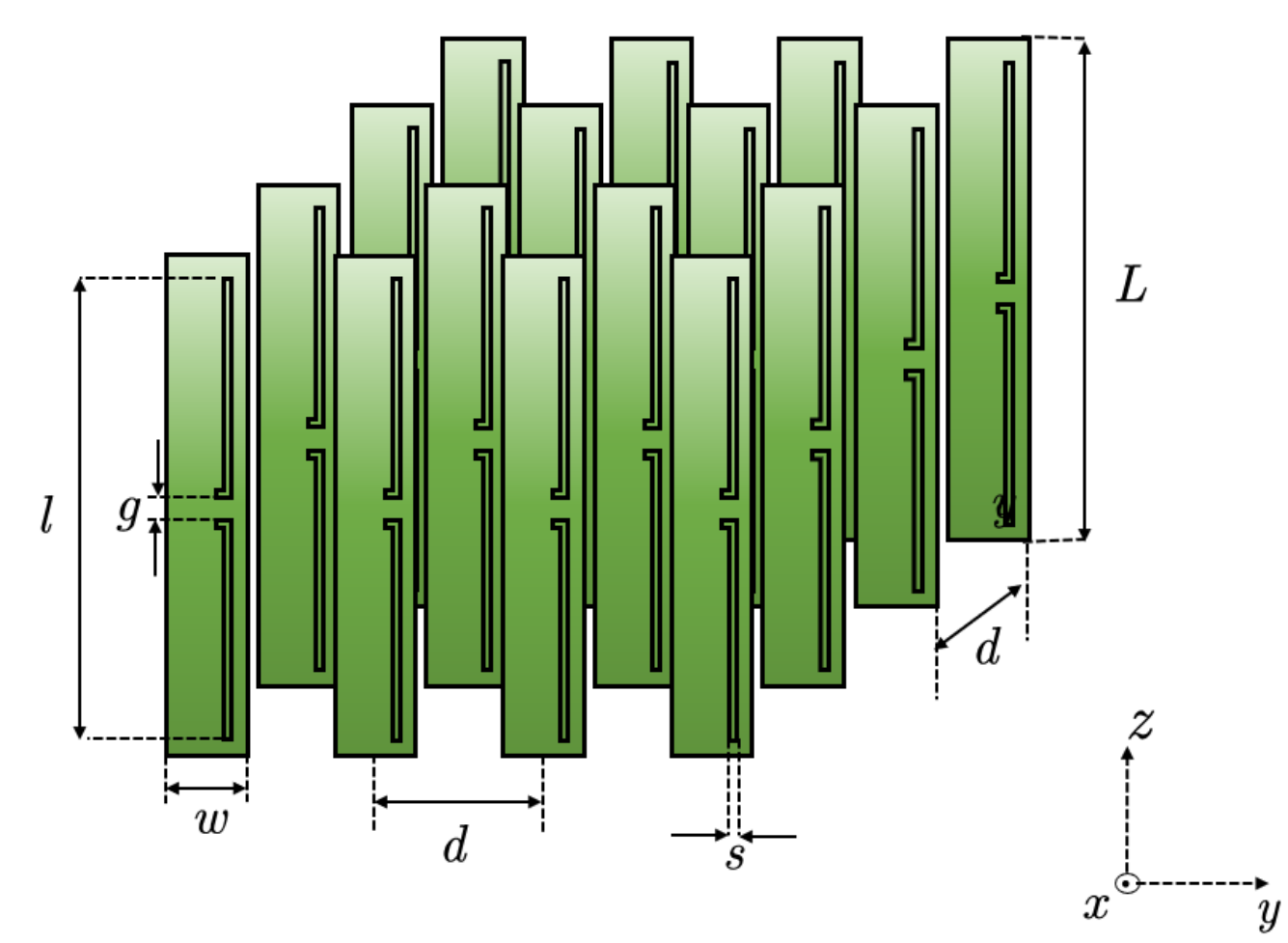}
    \caption{The antenna array of $4\times 4$ printed dipole antennas}
    \label{fig:planar_array}
\end{figure}

\begin{table*}[t]
    \centering
    \caption{The prediction accuracy of the proposed model based on different training methods.}
    \begin{tabularx}{\textwidth}{c|c|c|c}
       \hline
       The method of training model & 50 Epochs & 50 Epochs + Cosine LR with Warm-up & 100 Epochs + Cosine LR with Warm-up \\
       \hline
       The prediction accuracy of the proposed model & $95\%$ & $97.5\%$ & $98.5\%$ \\
        \hline
    \end{tabularx}
    \label{tab:V}
\end{table*}

\subsection{Data Acquisition and Model Training}
\label{sec:data_acquisition_planar}
Initially, we model the $4\times 4$ uniform printed dipole antenna array through the CST software, as illustrated in Fig. \ref{fig:planar_array}. The parameters of the planar antenna align with those of the four printed dipole array discussed in Sec. \ref{sec:numericalResult}.\par

In the $4\times 4$ uniform planar antenna array, the antenna spacing $d$ is randomly taken within the range of $\left[0.15\lambda, 0.50\lambda\right]$. A total 351 sets of uniform $4\times 4$ planar arrays with different antenna spacings were subject to simulation. Likewise, we sample the full electric field at $5^{\circ}$ intervals in the direction $\left(\theta,\phi\right)$. The excitation coefficients for each direction is subsequently determined through the application of Eq. (\ref{eq:6}). In total, $351\times 2664$ sets of the radiated electric field, along with their corresponding excitation coefficients for antenna arrays with varying spacings and orientations, have been obtained.\par

In Sec. \ref{sec:network}, the ``Network Architecture" specifies that the input and output data dimensions are $M\times 4\times 1$ and $M\times 2\times 1$, respectively. Here, $M$ is equal to 16 during the optimization of the excitation coefficients for the $4\times 4$ UPA.\par

Similar to the training procedure of the ``super-gain" prediction model discussed in Sec. \ref{sec:array_gain}, we compared the predictive performance of the model across different training scenarios. Specifically, the analysis involved evaluating the accuracy of the model after 50 epochs of training, 50 epochs of training with the warm up aided cosine LR schedule, and 100 epochs of training with the warm up aided cosine LR schedule. The results, illustrated in Table \ref{tab:V}, indicated that the prediction accuracy is approximately $95\%$ after 50 epochs of training. However, this accuracy improved to around $97.5\%$ when the warm up aided cosine LR schedule was incorporated. To further enhance the predictive capability, increasing the number of training epochs to 100 in conjunction with the the warm up aided cosine LR technology resulted in a prediction accuracy exceeding $98.5\%$. The results underscore the significant enhancement in prediction accuracy facilitated by the integration of the warm up aided cosine LR schedule into the training process of the model.\par

\subsection{Numerical results and analysis}
\label{sec:numerical_planar_array}
After undergoing 100 rounds of training along with the warm-up cosine-assisted method, the trained model was employed to forecast the excitation coefficients in different directions within the planar array. Table \ref{tab:VI} depicts the directivity of the $4\times 4$ printed dipole array in different directions with the antenna spacing of $0.25\lambda$ obtained by our proposed model and the coupling matrix-based method. The results of the simulation indicate a high degree of similarity between the directivity pattern obtained by the proposed model and the pattern derived from the coupling matrix-based method.\par

\begin{table}[h]
    \centering
    \caption{When the antenna spacing is $0.25\lambda$, the directivity obtained by different methods in various directions.}
    \begin{tabular}{c|c|c}
      \hline
       Direction $\left(\theta,\phi\right)$ & Coupling matrix-based method & \textbf{Proposed model} \\
      \hline
        $\left(45^{\circ},45^{\circ}\right)$ & 11.5 & \textbf{12.2}\\
      \hline
        $\left(65^{\circ},45^{\circ}\right)$ & 30.4  & \textbf{30.3}\\
      \hline
        $\left(70^{\circ},45^{\circ}\right)$ & 37.6 & \textbf{38.8}\\
      \hline
    \end{tabular}
    \label{tab:VI}
\end{table}

\section{Conclusions}
\label{sec:conclusion}
In this paper, we propose a MultiTransUNet-GAN model to predict the excitation coefficients to achieve the ``superdirectivity" and ``super-gain" in compact uniform linear and planar antenna arrays. Our proposed model integrates the generative adversarial mechanism, incorporating a multi-level guided attention module and a multi-scale skip connection within the generator network. Notably, the multi-level guided attention module combines a Transformer module and a GSA module to enhance the feature extraction capabilities of our model. During the model training, we incorporate the NMSE between the generated optimal excitation coefficients and the actual value into the objective function, and implement the warm up aided cosine learning rate scheduler to enhance the predictive accuracy of our model. We validated our method with a 4-element printed dipole uniform linear array and a $4\times 4$ printed dipole uniform planar array operating at 1.6 GHz. Our proposed model demonstrates a high level of prediction accuracy in compact uniform linear and planar arrays. Overall, our proposed model for predicting the excitation coefficients offers notable advantages over alternative approaches.\par

\bibliographystyle{IEEEtran}
\bibliography{ref}

\begin{IEEEbiography}
[{\includegraphics[width=1in,height=1.25in,clip,keepaspectratio]{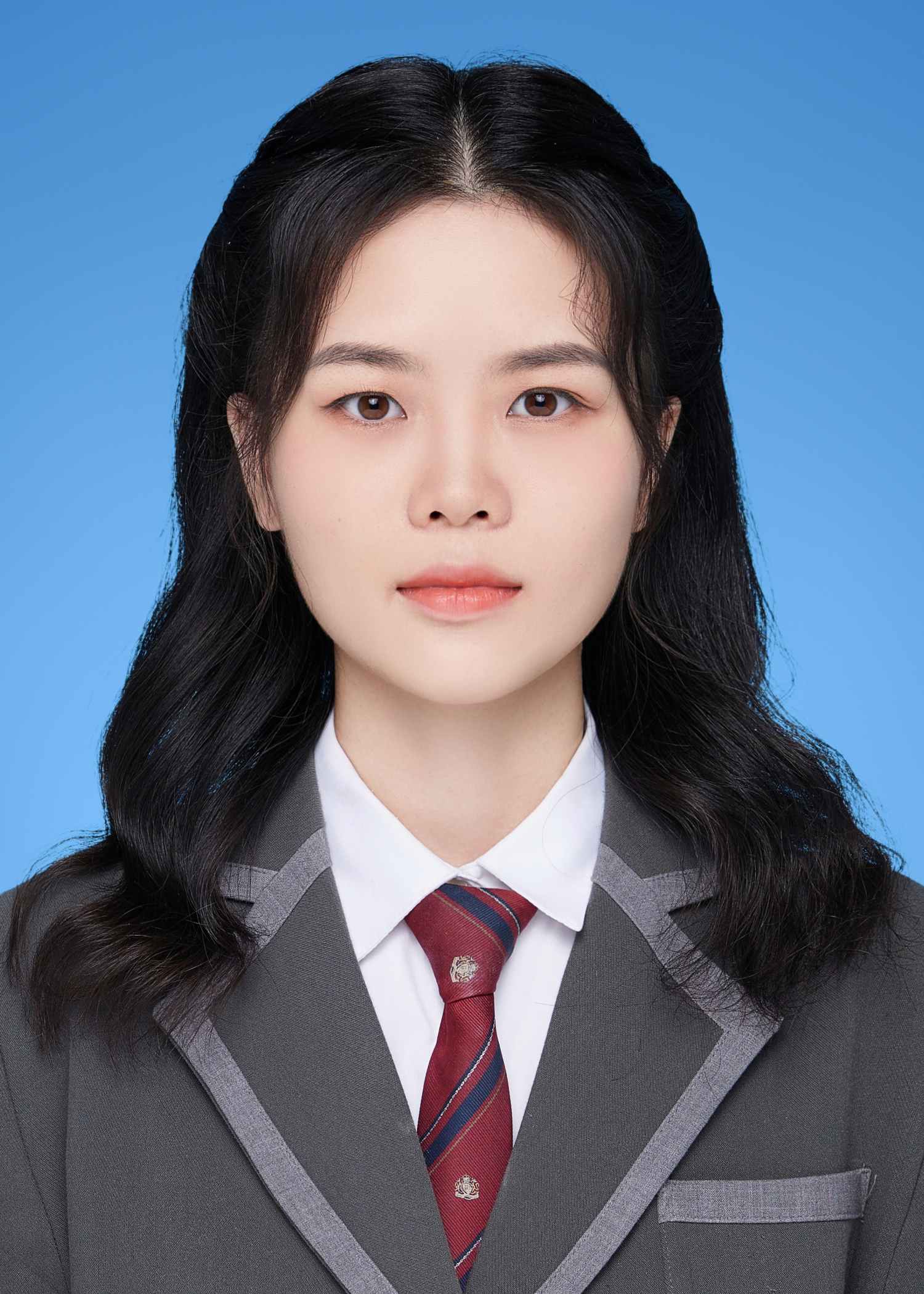}}]{Yali Zhang}
received the B.Sc. degree in Communication Engineering from the School of Electronic Information and Communications, Huazhong University of Science and Technology, Wuhan, China, in 2023. She is currently pursuing the M.Sc. degree with the School of Electronic Information and Communications from Huazhong University of Science and Technology, Wuhan, China. Her research interests include machine learning, wireless communications, array signal processing, and superdirective antennas.
\end{IEEEbiography}

\begin{IEEEbiography}[{\includegraphics[width=1in,height=1.25in,clip,keepaspectratio]{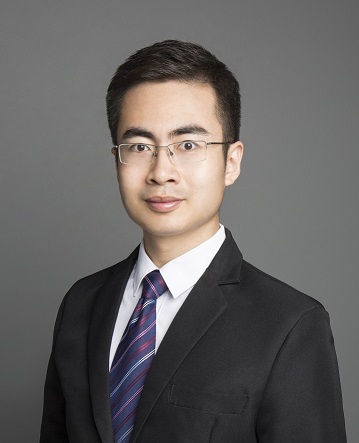}}]{Haifan Yin}
(Senior Member, IEEE) received the B.Sc. degree in electrical and electronic engineering and the M.Sc. degree in electronics and information engineering from the Huazhong University of Science and Technology, Wuhan, China, in 2009 and 2012, respectively, and the Ph.D. degree from Télécom ParisTech in 2015. From 2009 to 2011, he was a Research and Development Engineer with the Wuhan National Laboratory for Optoelectronics, Wuhan, working on the implementation of TD-LTE systems. From 2016 to 2017, he was a DSP Engineer at Sequans Communications (IoT chipmaker), Paris, France. From 2017 to 2019, he was a Senior Research Engineer working on 5G standardization at Shanghai Huawei Technologies Company Ltd., where he has made substantial contributions to 5G standards, particularly the 5G codebooks. Since May 2019, he has been a Full Professor with the School of Electronic Information and Communications, Huazhong University of Science and Technology. His current research interests include 5G and 6G networks, signal processing, machine learning, and massive MIMO systems. He was the National Champion of 2021 High Potential Innovation Prize awarded by the Chinese Academy of Engineering, a recipient of the China Youth May Fourth Medal (the top honor for young Chinese), and a recipient of the 2024 Stephen O. Rice Prize.
\end{IEEEbiography}

\begin{IEEEbiography}
[{\includegraphics[width=1in,height=1.25in,clip,keepaspectratio]{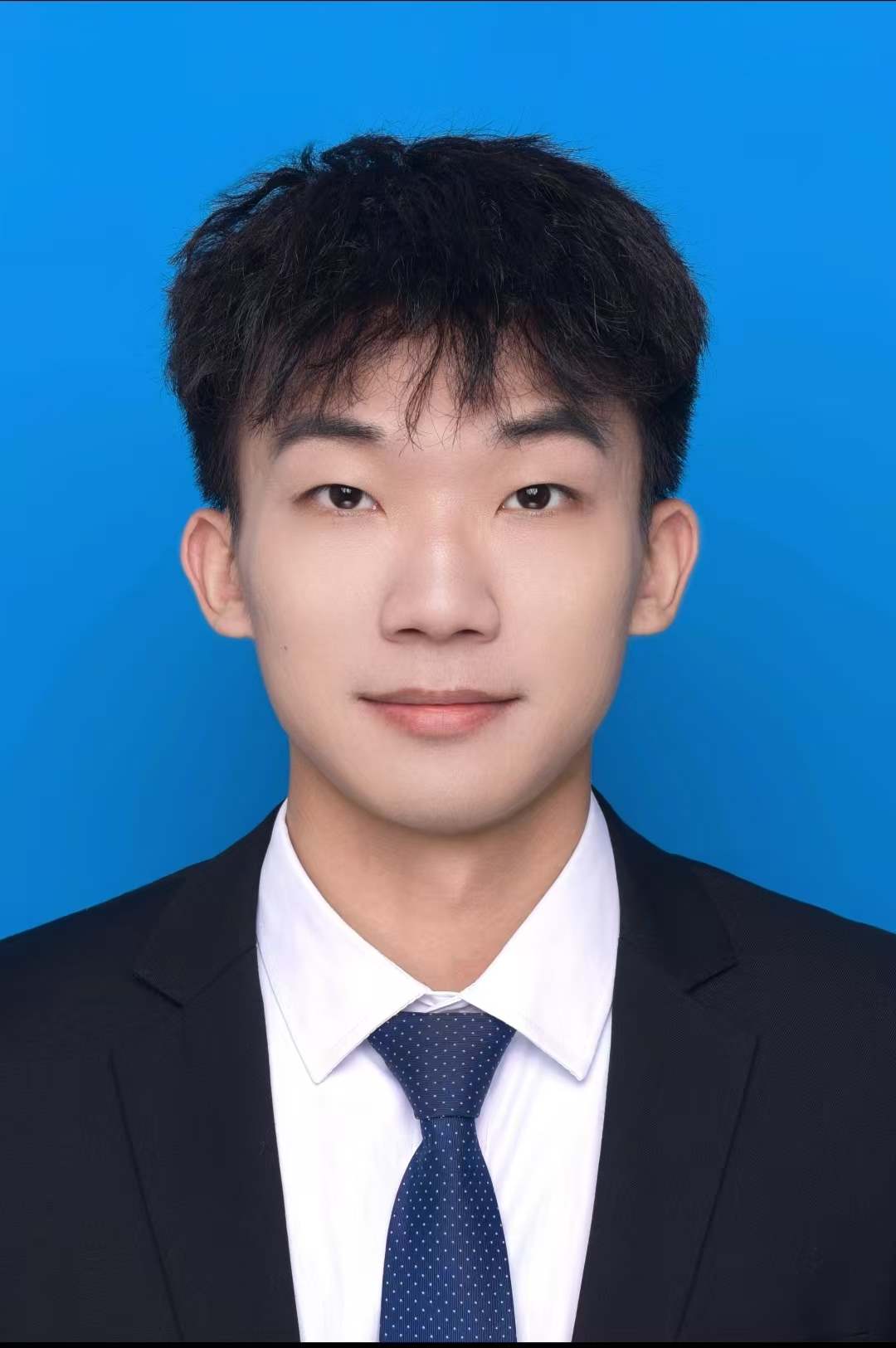}}]{Liangcheng Han}
received the B.Sc. degree from the School of Electronic Information and Communications at Huazhong University of Science and Technology in 2021. He then commenced his Master's studies at the same institution and transitioned to the Ph.D. program in 2023. He has been awarded the National Scholarship for Graduate Students, the National Gold Award in the Internet+ competition, and the First Prize in the National Challenge Cup competition. His research interests include wireless communications, array signal processing, and superdirective antennas.
\end{IEEEbiography}

\end{CJK}
\end{document}